\documentclass[preprint,sort&compress,numbers]{elsarticle}

\journal{Acta Astronautica}

\usepackage{amsmath}
\usepackage{amssymb}
\usepackage{subcaption}

\newcommand{\norm}[1]{\left\Vert#1\right\Vert}
\newcommand{\abs}[1]{\left\vert#1\right\vert}

\DeclareMathAlphabet{\mbf}{OT1}{ptm}{b}{n}

\newcommand{\mbfhat}[1]{{\hat{\mbf{#1}}}}

\newcommand{\p}{\partial}
\newcommand{\ura}[1]{{\underrightarrow{{#1}}}}

\newcommand{\trans}{{\ensuremath{\mathsf{T}}}}

\begin{document}

\begin{frontmatter}

\title{Autonomous Station Keeping of Satellites in Areostationary Mars Orbit: A Predictive Control Approach}

\author[label1]{Robert D. Halverson}

\author[label2]{Avishai Weiss}

\author[label1]{Gabriel Lundin}

\author[label1]{Ryan J. Caverly}

\affiliation[label1]{organization={Department of Aerospace Engineering \& Mechanics, University of Minnesota},
            city={Minneapolis}, 
            state={MN},
            postcode={55455},
            country={United States}}
            
\affiliation[label2]{organization={Mitsubishi Electric Research Laboratories},
            city={Cambridge},
            state={MA},
            postcode={02139}, 
            country={United States}}

\begin{abstract}

The continued exploration of Mars will require a greater number of in-space assets to aid interplanetary communications. Future missions to the surface of Mars may be augmented with stationary satellites that remain overhead at all times as a means of sending data back to Earth from fixed antennae on the surface. These areostationary satellites will experience several important disturbances that push and pull the spacecraft off of its desired orbit. Thus, a station-keeping control strategy must be put into place to ensure the satellite remains overhead while minimizing the fuel required to elongate mission lifetime. This paper develops a model predictive control policy for areostationary station keeping that exploits knowledge of non-Keplerian perturbations in order to minimize the required annual station-keeping $\Delta v$. The station-keeping policy is applied to a satellite placed at various longitudes, and simulations are performed for an example mission at a longitude of a potential future crewed landing site. Through careful tuning of the controller constraints, and proper placement of the satellite at stable longitudes, the annual station-keeping $\Delta v$ can be reduced relative to a na\"ive mission design.

\end{abstract}


\end{frontmatter}

\section{Introduction}

As the frequency of visits to Mars---robotic or crewed---is expected to grow in the coming decades, there will exist a greater need for reliable communication between Earth and the surface of Mars. A means of supporting this communication link may be via satellites in orbit around Mars---in particular, ones that remain overhead at all times. Stationary satellites have been used for decades around Earth for the purpose of general telecommunications (e.g., television, inter-satellite communication), weather \& surface imaging, and defense applications. An areostationary Mars orbit (AMO) is the Martian equivalent to a geostationary Earth orbit (GEO). These are circular, equatorial orbits with a period matching the rotation of Mars such that the satellite appears fixed in the sky to a ground observer. Areostationary satellites can deliver greater communication availability between Earth and the surface of Mars to support an expanding presence on the surface. The remainder of this introduction includes a review of literature surrounding motivation for areostationary satellites and previous work regarding station-keeping strategies. Section~\ref{sec:contributions} will discuss the contributions of this paper and the research explored within.

\subsection{Past, Present, and Proposed Mars Exploration}
There exist several proposals describing the need for areostationary satellites to be placed around Mars. NASA has specifically recommended these orbiters be prioritized for study, use, and development in the next decade~\cite{Montabone2021}. An early proposal cites an areostationary orbiter as ``unsurpassable'' with regards to continuous coverage and high levels of bandwidth and connectivity, which led to the design of the Mars Areostationary Relay Satellite (`MARSAT'), a proposed satellite that would serve as part of a telecommunications infrastructure in support of Mars exploration~\cite{Edwards2001}. A Mars Global Navigation Satellite System has also been proposed, where a constellation of satellites would enable real-time navigation on the surface of Mars~\cite{FernandezGNSS}. The proposed constellation includes fifteen satellites, three of which are stationary. Constellations of small areostationary satellites, such as those proposed in Ref.~\cite{Lock2016}, may provide a network of satellites continually available overhead to support humans and vehicles in the process of establishing a more consistent presence on the surface, while ensuring continuous access to high data return to Earth. Other data return concepts commonly require satellites in AMO, including relay orbiters to provide continuous high-bandwidth coverage to the surface and other satellites in a lower orbit back to Earth~\cite{Edwards2016,Breidenthal2018,Babuscia2017,Hastrup2003}. The requirements for orbital determination of areostationary relay satellites has also been studied, while analyzing the non-Keplerian perturbations that move satellites off from their desired orbit~\cite{Romero2017}. 

Several missions have been deployed to support ground or low orbit operations for different purposes. The Mars Atmosphere and Volatile EvolutioN (`MAVEN') spacecraft has been used to observe the Martian atmosphere and surrounding space weather since 2014, having parked in several different orbits to measure the upper atmosphere, ionosphere, solar winds, and interplanetary magnetic field conditions~\cite{Lee2017,Jakosky2015}. MAVEN and other missions have supported the research into wind modeling and surface atmospheric conditions, which are important qualities to take into account for future missions to the surface~\cite{guzewich2020measuring,Lewis1995}. Notably, there exist massive dust storms on Mars, which could carry extreme consequences for humans on the surface. Future capabilities have been proposed that rely on areostationary satellites to make continuous observations over a wide area, filling gaps left by other missions regarding the forecasting of dust storms and other atmospheric weather phenomena~\cite{Montabone2018}. Aside from the solar and atmospheric environment, the gravitational field of Mars has been studied through missions such as the Mars Global Surveyor (MGS), Mars Reconnaissance Orbiter (MRO), and Mars Odyssey (ODY). These satellites have provided the information necessary to develop high-fidelity harmonic models of Mars' gravitational field~\cite{Lemoine2001,Genova2016}. Data from areostationary satellites could supplement development of even more accurate gravitational models. 

Satellites in AMO have a direct line of sight to Earth for a majority of time, dependent on the geometry and relative positions of the Earth, Mars, and Sun. A worst-case estimate is that a direct line of sight is available 94.7\% of the time. Details on calculations to support this estimate are found in Appendix A.

\subsection{Station Keeping \& Control Strategies}
Station keeping is required to keep AMO satellites at a specified longitude due to the presence of non-Keplerian perturbations, which effectively push and pull the satellites off of their nominal trajectory~\cite{Alvarellos2010,Konopliv2006}. Station keeping is generally achieved via small thrusts using chemical or electric propulsion. While these immediate injections of linear momentum are usually small, the total fuel used over the lifetime of a satellite can become significant. Thus, the operational lifetime of these satellites is limited by the amount of fuel they can carry, and as such, a controller should be developed that is capable of minimizing the magnitude, duration, and frequency of thrusts. A method to quantify the fuel required for orbital maneuvers is via a change in velocity, $\Delta v$, which is typically computed as the sum of all instantaneous impulses normalized by satellite mass. In low-thrust applications, thrust is often applied in a more continuous fashion. Thus, the computation of $\Delta v$ in these applications is modified to be the integral of the thrust over time normalized by the satellite's mass. By minimizing $\Delta v$, fuel consumption is directly minimized, which in turn extends the useful life of the satellite, increases payload mass fraction, and/or reduces cost.

Stationary satellites are controlled to maintain their position within a prescribed station-keeping window. This can be thought of as an allowable deadband such that the satellite doesn't exactly maintain its position over a specific longitude and instead can drift slightly from its nominal circular orbit. The size of the station-keeping window for Earth applications is usually chosen to be small enough to prevent collisions between satellites. Relaxing the stationary constraint to some allowable drift window can provide significant savings regarding $\Delta v$ cost~\cite{Halverson2023}. In GEO, station keeping is often performed offline, where a ground station will send commands when required to maintain the satellite's position within station-keeping window constraints~\cite{soop1994handbook}. Time delay in communication between Earth and Mars varies between 8--22~minutes, making it difficult to command satellites in a Martian orbit from Earth.

Several implementations of autonomous station-keeping strategies have been proposed in recent years for GEO satellites, including fuzzy logic and optimal control~\cite{Maghsoudi2019}, convex optimization methods~\cite{deBruijn2016}, and model predictive control (MPC)~\cite{Weiss2018}. MPC has the ability to utilize the predicted disturbances as well as knowledge of station-keeping window constraints to directly determine discrete thrust inputs over a receding prediction horizon. The MPC in Ref.~\cite{Weiss2018} achieves fuel-efficient station keeping via numerical optimization that penalizes the control effort over a finite horizon. There are several more examples of applying MPC to complex station-keeping problems, including split-horizon MPC with different prediction timescales for along-track and cross-track motion, on-off thruster quantization, and control allocation of satellites with boom assemblies~\cite{Caverly2020,Caverly2020_IEEE}. More comprehensive overviews on the use of MPC for spacecraft and other aerospace applications are found in Refs.~\cite{di2018real,petersen2023safe,eren2017model}.

A viable propulsion scheme for station keeping is the use of electric propulsion technology as an alternative to traditional chemical propulsion systems~\cite{Martinez-Sanchez1998}. Electric propulsion offers a significantly more efficient propulsive method compared to chemical propulsion in terms of specific impulse (ISP)~\cite{lev2019technological}. Applications of electric propulsion with an MPC policy have been investigated for GEO station keeping, including a complete control problem with station keeping, attitude control, and momentum management via MPC~\cite{Caverly2020_IEEE}. 

Recent work has been introduced to explore the deployment of areostationary satellites as a part of a larger Mars constellation. Ref.~\cite{pontani2022mars} describes the development of a nonlinear orbit control strategy for low-thrust deployment of satellites from a higher, more eccentric orbit into their required position in the constellation, which includes three AMO satellites placed equidistant around the planet. The work in~\cite{pontani2022mars} is especially useful in further supporting the proposition and viability of using electric propulsion systems for station keeping of AMO satellites, and characterizes the usefulness of augmenting a constellation with stationary satellites.

\subsection{Station Keeping on Areostationary Mars Orbit}
Previous work has aimed to quantify the $\Delta v$ required for an areostationary satellite through an investigation into the disturbance forces present around Mars~\cite{Silva2013}. Importantly, there exist gravitational perturbations due to Mars being a non-spherical central body, which creates a non-homogeneous gravitational field~\cite{Alvarellos2010,Vallado2004}. Further, investigations into the stability of areostationary longitude placements and periodic orbits around those points have taken place in order to better characterize the gravitational field~\cite{Liu2012}. A study of these perturbations is usually sufficient for an introductory investigation into the $\Delta v$ required for station keeping over an entire year. However, a simple quantification of fuel required without implementing a realistic control scheme overlooks complexities that may arise due to nonlinear coupling of disturbances and their affect on orbital motion. To overcome this gap, use of an implementable control strategy within a high-fidelity simulation can provide a mission designer a much more accurate insight into the true $\Delta v$ required.

Research into the implementation of areostationary satellites has included discussing the feasibility and full extent of the coverage of Mars via satellites in AMO~\cite{Colella2017}. The perturbations that cause large relative longitudinal and radial motion were found to be useful to mission design via proper satellite placement. With minimal corrective maneuvers, the total $\Delta v$ required would be $60$~m/s per year, and without station keeping, loss of coverage would occur in roughly one month. These results were derived from a simplified gravitational model, neglecting higher-order spherical harmonics and the effect of the Sun. Other research has been performed to investigate the viable control schemes for AMO satellite station keeping, including computation of solar eclipse temporal patterns for planning of station-keeping strategies to meet constraints~\cite{Romero2014}, maintenance of a zero-inclination orbit~\cite{Romero2015}, or use of a distributed internal control feedback strategy for coordination of areostationary satellite constellations~\cite{Sin2020}. Notably, the control strategy in Ref.~\cite{Sin2020} focused entirely on satellite orbital distancing rather than maintaining position over a specified longitude, without a minimum-fuel objective. There has yet to be an implementable autonomous station-keeping control strategy proposed and verified in a high-fidelity simulation that provides an expectation of fuel required for a full year of station keeping over specific longitudes. 

\subsection{Contributions}\label{sec:contributions}
This paper aims to fill the gap in the literature by describing a fuel-optimized and constraint-oriented station-keeping control strategy via MPC. Previous work included the application of linear quadratic MPC to a satellite in a stable areostationary orbit, and compared its results and fuel required to na\"{\i}ve station-keeping policies~\cite{Halverson2021}. Those results were expanded to describe certain anomalies in the control law and its interaction with the environment, and new results utilizing nonlinear MPC were introduced to further decrease the annual $\Delta v$ required for areostationary station keeping~\cite{Halverson2023}. In this work, these preliminary studies are expanded upon by providing additional details on the AMO environment and introducing an updated predictive control formulation. Further, the effect of $\Delta v$ requirement vs. mission start epoch is explored, while also investigating modifications to important tuning parameters in the MPC policy.

The novel contributions of this work include: 
\begin{enumerate}
    \item Implementing a predictive control policy to an AMO satellite that directly optimizes $\Delta v$;
    \item Providing a benchmark on $\Delta v$ required for station keeping at all longitudes around Mars;
    \item Investigating and discussing the tuning parameters and design considerations for placing a satellite in AMO above a location of potential future scientific interest.
\end{enumerate}

A thorough description of the environmental model used in simulation is introduced in Section~\ref{Sec:EnvModel}, including all important perturbations. The control law is then discussed in Section~\ref{sec:MPC}, including modifications of the nonlinear MPC policy from Ref.~\cite{Halverson2023}, which aims to directly minimize the required annual $\Delta v$. In Section~\ref{sec:longSweep}, the nonlinear MPC policy is implemented in an exhaustive sweep of longitudes such that a benchmark of the required annual station-keeping $\Delta v$ is provided. Further, we explore the implementation in a mission design of an areostationary satellite above a location of scientific interest and a potential future crewed landing site---the Southern Meridiani Planum~\cite{clarke2017southern}---in Section~\ref{sec:SMP}. Important design considerations in this case are discussed, and an extension to applying station keeping at a stable longitude is introduced to further reduce fuel cost in Section~\ref{sec:stableLong}.

\section{Preliminaries}

\subsection{Notation}
The following notation is used throughout this paper. $\mathcal{F}_a$ is a reference frame defined by a set of three orthonormal dextral basis vectors, $\underrightarrow{a}^1$, $\underrightarrow{a}^2$, and $\underrightarrow{a}^3$. The physical vector describing the position of point $p$ relative to point $q$ is $\underrightarrow{r}^{pq}$. Similarly, the angular velocity of $\mathcal{F}_b$ relative to $\mathcal{F}_a$ is $\underrightarrow{\omega}^{ba}$. The mapping between a physical vector resolved in different reference frames is given by a direction cosine matrix (DCM) $\mbf{C}_{ba}$, where $\mbf{C}_{ba} \in \mathbb{R}^{3\times3}$, $\mbf{C}_{ba}\mbf{C}_{ba}^\trans = \mbf{1}$, and det($\mbf{C}_{ba}$) = +1, where $\mbf{1}$ is the identity matrix. For example, $\mbf{u}_b = \mbf{C}_{ba}\mbf{u}_a$, where $\mbf{u}_a$ is $\underrightarrow{u}$ expressed in $\mathcal{F}_a$, $\mbf{u}_b$ is $\underrightarrow{u}$ expressed in $\mathcal{F}_b$, and $\mbf{C}_{ba}$ represents the attitude of $\mathcal{F}_b$ relative to $\mathcal{F}_a$. The DCM $\mbf{C}_{i}(\theta)$ is a principle rotation around the $i^{\textrm{th}}$ basis vector by the angle $\theta$.

For discrete-time systems, a subscript notation of $k|t$ is used, denoted as a state or control input at $k$ time steps ahead of the current time $t$ (i.e., $\mbf{x}_{k|t}$ is read as the predicted state at $k$ time steps ahead of current time $t$).

\subsection{Important Reference Frames}
Three reference frames are defined. A Mars-centered inertial (MCI) frame $\mathcal{F}_a$, similar to an Earth-centered inertial (ECI) frame, is defined with $\underrightarrow{a}^1$ pointing outwards from the center of Mars towards $0^\circ$ latitude and longitude at J2000, and $\underrightarrow{a}^3$ pointing North, perpendicular to equatorial plane. Finally, $\underrightarrow{a}^2$ completes the orthonormal, dextral reference frame. For completeness, J2000 is defined as January 1st, 2000 at noon GMT Earth-time. Zero longitude on Mars (and that of all other important celestial bodies) is defined by the International Astronomical Union (IAU) through careful recommendations on values for the direction of the north pole of rotation and the prime meridian~\cite{abalakin2002report}.

Hill's frame $\mathcal{F}_h$ is a rotating frame, where $\underrightarrow{h}^1$ lies in the equatorial plane and points towards a specified longitude from the center of Mars, $\lambda$, $\underrightarrow{h}^2$ points in the along-track direction of the orbit, and $\underrightarrow{h}^3$ points north, or in the cross-track direction.  The DCM describing the attitude of $\mathcal{F}_h$ relative to $\mathcal{F}_a$ at time $t$ after J2000 is given by $\mbf{C}_{ha} = \mbf{C}_3(\lambda+nt)$, where $\lambda$ is the desired station-keeping longitude and $n$ is the orbit mean motion. 

\begin{figure}[t!]
	\includegraphics[width=0.95\linewidth]{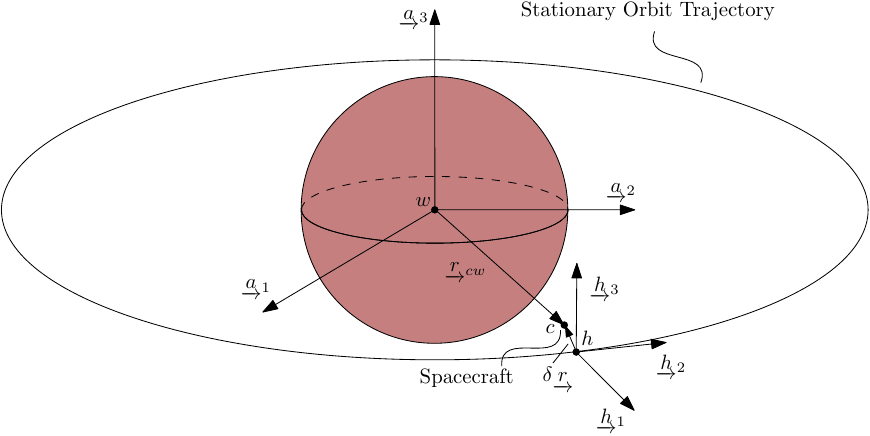} 
	\centering
	\caption{Illustration of the MCI frame (aligned with $\mathcal{F}_a$) and Hill's frame ($\mathcal{F}_h$) attached to the spacecraft (point $c$) in a circular equatorial areostationary orbit. This figure also shows the spacecraft drift, $\delta \ura{r}$, relative to the station-keeping window fixed in Hill's frame and centered at point $h$. Stationary orbit and planet sizes are not to scale. }\label{fig:hillsFrame}
\end{figure}

The Mars-Centered-Mars-Fixed frame (MCMF) $\mathcal{F}_b$ is similar to an Earth-Centered-Earth-Fixed frame (ECEF), and is defined relative to MCI in a similar manner as $\mathcal{F}_h$. In this case, the DCM describing this rotation is given by $\mbf{C}_{ba} = \mbf{C}_3(\lambda + \omega t)$, where $\omega$ is the nominal rotation rate of Mars. For the sake of propagating the MCMF frame, we assume that Mars' rotation rate is the same at each epoch, and the full duration of each study herein. Given a stationary orbit configuration and the assumptions described, $\mathcal{F}_h$ and $\mathcal{F}_b$ are aligned for all $t$. See Figure~\ref{fig:hillsFrame} for an illustration of these frames.

\subsection{Assumptions and Objectives}
The control objective of this work is to perform station keeping with minimal fuel usage. Proper station keeping is achieved when the satellite remains within a prescribed station-keeping window, where a limit of drift in both latitude and longitude is specified. The satellite's control system is allowed (and encouraged) to maximize drift within the station-keeping window such that the $\Delta v$ required is minimized.

It is assumed that the satellite is initially in a stationary orbit with zero eccentricity and an orbital period equal to the rotational period of Mars. This work only considers the translational dynamics of the satellite, and thus the spacecraft is modeled as a point-mass. Furthermore, it is assumed that the satellite is continually able to thrust in any positive or negative orthogonal direction in $\mathcal{F}_h$. This can be interpreted as the satellite having six thrusters mounted in a fashion such that the satellite can generate a thrust force in any orthogonal direction at any time (i.e., radially, cross-track, and along-track). It is assumed that all thrusters are throttle-able such that they may produce any thrust between zero and their maximum thrust. A real implementation would include control allocation according to the satellite hardware, which---according to a particular thruster configuration---would achieve the required $\Delta v$ impulse (i.e., this work aims to provide the required $\Delta v$ for a satellite in AMO, regardless of thruster configuration). The mass of the satellite is assumed to be constant, which may introduce conservatism in the thrust magnitude required for station keeping. Regardless, because the parameter we aim to minimize is $\Delta v$ (thrust impulse normalized by mass), the $\Delta v$ required for each result is not affected by this assumption. Finally, this work neglects the occultation of the Sun by Mars in solar radiation pressure calculations. 

To improve the computational efficiency of the simulations, the positions of Mars' moons relative to Mars are calculated assuming that the bodies maintain constant orbital elements for the duration of the simulation. Likewise, the orbits of Mars and Jupiter around the Sun are assumed to have constant orbital elements over the course of one simulation (roughly one orbital period of Mars around the Sun). These approximations are deemed acceptable, as the ephemeris data provided by JPL Horizons~\cite{JPLHorizons} confirms that less than $1\%$ error in the position vectors is incurred through this assumption.

\section{Environment Model} \label{Sec:EnvModel}
A high-fidelity environmental model including gravitational and other perturbation forces is developed to investigate the station keeping problem. The purpose is two-fold; first, to provide an accurate prediction for satellite drift that is used in the predictive control policy, and second, to ensure this investigation provides a proper estimate of the station-keeping control required for satellites in AMO. 

The satellite model is given by Newton's law of gravitation augmented with non-Keplerian perturbations. Gravitational disturbances influence a stationary satellite to drift from its desired location on orbit. Specifically, we take into account the presence of the Sun, Jupiter, and Mars' two moons (Phobos and Deimos) as the celestial bodies that cause the most significant disturbances. Further, we consider solar radiation pressure caused by the transfer of momentum between photons from the Sun and the satellite. Lastly, the largest driver in a satellite's drift in the Martian environment is the gravitational spherical harmonics due to the planet's non-homogeneous distribution of mass. A complete 3-degree-of-freedom simulation of this environment is modeled using the equations of motion
\begin{equation}
\label{eq:EoM1}
\ddot{\mbf{r}}_a^{cw} = -\mu \frac{\mbf{r}_a^{cw}}{\norm{\mbf{r}_a^{cw}}^3} + \mbf{a}^p_a + \frac{1}{m_B}\mbf{C}_{ha}^\trans \mbf{f}_h^\text{thrust},
\end{equation}
where $\mbf{r}_a^{cw}$ is the position vector of a point $c$ at the spacecraft center of mass  relative to a stationary point $w$ at Mars' center expressed in $\mathcal{F}_a$;  $\mu = 42828.37$~km$^3$/s$^2$ is the standard gravitational parameter of Mars; $\mbf{a}^p_a$ contains the perturbing accelerations acting on the satellite expressed in $\mathcal{F}_a$; $m_B$ is the mass of the spacecraft; and $\mbf{f}_h^\text{thrust}$ is the thrust force applied to the spacecraft expressed in $\mathcal{F}_h$. The perturbations that are included in simulation are described in the following sub-sections. 

We neglect minor perturbations from disturbances such as tides of solid mass or CO$_2$, relativistic effects, radiation pressure emitted from Mars, gravitational forces from other planets, and atmospheric drag~\cite{chao2018,MarsGRAM}. These perturbations do not produce meaningful disturbances to areostationary satellites over the expected lifetime of the satellite and thus do not require compensation.

\subsection{Gravitational Spherical Harmonics}\label{sec:spherHarm}
The most prevalent perturbation in the Martian orbital environment comes from the non-homogeneous gravitational field formed by the non-homogeneous distribution of mass within the planet. A well-known gravitational disturbance caused by this effect is the J2 perturbation, which is a driving factor in the precession of inclined orbits. A realistic mathematical model that captures the non-spherical nature of Mars is used, specifically taking into account higher-order ($n>2$) terms in the aspherical gravitational potential function ($\Phi$)~\cite{Vallado2004}. These higher-order terms become important in the Mars gravitational environment due in part to the size of the large mountains and valleys on Mars relative to its small planetary radius, as well as ice deposits and many other factors that contribute to a generally uneven distribution of mass within the planet. 

An infinite series potential function of a spheroid is used to define the gravitational potential of a primary body. This potential function is calculated first as a function of satellite latitude ($\phi$), longitude ($\lambda$), and distance from the center of Mars ($r$), given by~\cite{Vallado2004,Alvarellos2010,DeRuiter2013}
\begin{multline}\label{eq:gravPotential}
\Phi(r,\lambda,\phi) = \frac{\mu_\text{M}}{r} \left[ -\sum_{n = 2}^{\infty}J_n \left( \frac{R_\text{M}}{r} \right)^n P_n(\sin\phi) \right. \\ \left.
+ \sum_{n = 2}^{\infty} \sum_{m = 1}^{n} \left(\frac{R_\text{M}}{r}\right)^n P_{n,m}(\sin\phi) \left[ C_{n,m}\cos(m\lambda) + S_{n,m}\sin(m\lambda)   \right]	\right],
\end{multline}
where $\mu_\text{M}$ is the Mars gravitational constant ($GM$); $R_\text{M}$ is the planetary radius of Mars; $J_n$ is the coefficient that defines the zonal harmonics; $C_{n,m}$ and $S_{n,m}$ are coefficients for the remaining spherical harmonics, specifically sectorial and tesseral harmonics; and $P_{n,m}$ are the associated Legendre polynomials. At $0^\textrm{th}$ order,~\eqref{eq:gravPotential} simplifies to $\mu/r$, which leads to the two-body Keplerian equations of motion. In-depth information regarding the gravitational potential function, means of finding the gravitational coefficients, and definitions of the associated Legendre polynomials can be found in Ref.~\cite{Vallado2004}. 

The different types of spherical harmonics are described as zonal, sectorial, and tesseral harmonics. The zonal harmonics reflect the primary body's oblateness, and can be described as having greater mass distribution around bands of latitudinal regions of the planet. Sectorial harmonics take into account differences in mass distribution in longitudinal regions (similar to the vertical stripes on a beach ball). The tesseral harmonics attempt to model specific regions on the planet that depart from a perfect sphere, and can be thought of a checkerboard pattern, or tiles of high or low gravitational attraction. These types of spherical harmonics are visualized graphically in Ref.~\cite{chao2018}.  Taking into account these three types of spherical harmonics allow for development of a high fidelity model of Mars' gravitational field. 

The gravitational acceleration, which takes into account the spherical harmonics that come about from coefficients in the potential function in~\eqref{eq:gravPotential}, can be expressed in $\mathcal{F}_a$ as
\begin{equation*}
	\mbf{a}_a^{\text{spher}} = \vec{\nabla}\Phi(x_a,y_a,z_a) = \frac{\p}{\p x_a} \Phi\ura{a}^1 + \frac{\p}{\p y_a} \Phi\ura{a}^2 + \frac{\p}{\p z_a} \Phi\ura{a}^3,
\end{equation*}
where the gravitational potential function variables $r$, $\delta$, and $\lambda$ first must be converted to Cartesian coordinates before a gradient is taken with respect to the Cartesian coordinates in $\mathcal{F}_a$ (i.e., $\Phi(r,\lambda,\phi) \Longrightarrow \Phi(x_a,y_a,z_a)$). This is achieved by a change of variables defined by 
\begin{equation*}
x=r\cos\phi\cos\lambda, \hspace{15pt} y=r\cos\phi\sin\lambda, \hspace{15pt} z=r\sin\phi,
\end{equation*}
where $r=\sqrt{x^2 + y^2 + z^2}$. From here, the necessary trigonometric identities can be used to complete the coordinate transformation. 

The definition of the perturbation due to the gravitational spherical harmonics that include higher-order terms from the nominal Keplerian orbit can be written as
\begin{align*}
\mbf{a}_a^\textrm{grav} = \mbf{a}_a^\textrm{spher} - \mu \frac{\mbf{r}_a^{cw}}{\norm{\mbf{r}_a^{cw}}^3},
\end{align*}
which removes the $0^\textrm{th}$ order terms from~\eqref{eq:gravPotential}, and recovers solely the disturbance force per unit mass.

Table~\ref{table:spherHarm_force} defines the gravitational spherical harmonics disturbance acting on the satellite at various longitudes. Data was collected from the Goddard Mars Model 2B (GMM-2B)~\cite{Lemoine2001} using MATLAB's \texttt{gravitysphericalharmonic.m} function. At each longitude, there is no significant change in disturbance force between orders $n=20$ and the maximum order available, $n=80$. For the simulations in this work, an order $n=5$ is selected, given that the disturbance force converges to very little practical change at order $n>5$, with a great amount of trade-off in computation time.

\begin{table} [t!]
	\begin{subtable}[]{1\textwidth}
		\center
		\begin{tabular}{l|c|c|c}  \hline 	
			Order & $a_{h1}^\textrm{grav}$ & $a_{h2}^\textrm{grav}$ & $a_{h3s}^\textrm{grav}$ \\  \hline \hline
		$2$  & $-6.721\times10^{-6}$ & $-9.976\times10^{-8}$ &  $8.814\times10^{-13}$  \\ 	
		$5$  & $-6.745\times10^{-6}$ & $7.932\times10^{-9}$ &  $-2.013\times10^{-8}$  \\
		$20$ & $-6.745\times10^{-6}$ & $8.002\times10^{-9}$ &  $-2.011\times10^{-8}$  \\
		\hline
		\end{tabular}
		\caption{Stable Longitude}
	\end{subtable}
	\begin{subtable}[]{1\textwidth}
		\center
		\vspace{5pt}
		\begin{tabular}{l|c|c|c} \hline 	
			Order & $a_{h1}^\textrm{grav}$ & $a_{h2}^\textrm{grav}$ & $a_{h3}^\textrm{grav}$ \\  \hline \hline
		$2$  & $-9.653\times10^{-6}$ & $-6.085\times10^{-7}$ &  $1.819\times10^{-12}$  \\ 	
		$5$  & $-9.493\times10^{-6}$ & $-5.039\times10^{-7}$ &  $5.645\times10^{-8}$  \\
		$20$ & $-9.493\times10^{-6}$ & $-5.040\times10^{-7}$ &  $5.642\times10^{-8}$  \\
		\hline
		\end{tabular}
		\caption{Unstable Longitude}
	\end{subtable}
	\begin{subtable}[]{1\textwidth}
		\center
		\vspace{5pt}
		\begin{tabular}{l|c|c|c}  \hline
			Order & $a_{h1}^\textrm{grav}$ & $a_{h2}^\textrm{grav}$ & $a_{h3}^\textrm{grav}$ \\  \hline \hline
		$2$  & $-8.451\times10^{-6}$ & $1.071\times10^{-6}$ &  $-2.292\times10^{-12}$  \\ 	
		$5$  & $-8.416\times10^{-6}$ & $1.204\times10^{-6}$ &  $2.767\times10^{-8}$  \\
		$20$ & $-8.416\times10^{-6}$ & $1.204\times10^{-6}$ &  $2.770\times10^{-8}$  \\
		\hline
		\end{tabular}
		\caption{Worst-Case Longitude}
	\end{subtable}
	\caption{Gravitational spherical harmonic disturbance acceleration components resolved in $\mathcal{F}_h$ at (a) $17.92^\circ$W, (b) $92^\circ$E, and (c) $148^\circ$W, varying with order of the spherical harmonic potential outlined in Function~\ref{eq:gravPotential}. Each disturbance acceleration has units of m/s$^2$ and is rounded to the third decimal place.} \label{table:spherHarm_force}
\end{table}

\subsection{Solar Radiation Pressure}
The influx of photons radiated from the Sun induces a pressure on the satellite known as solar radiation pressure (SRP). An expression for this perturbing acceleration is given by Equation (10c) in Ref.~\cite{Weiss2018}
\begin{equation*}
\mbf{a}^{\text{srp}}_a = C_{\text{srp}} S_{\text{facing}} \left[ \left( \frac{1 + c_{\text{refl}}}{2 m_B} \right) \left(  \frac{\mbf{r}_a^{cs}}{\norm{\mbf{r}_a^{cs}}} \right) \right], 
\end{equation*}
where $s$ denotes the center of the Sun, $C_{\text{srp}}$ is the solar radiation pressure constant, $S_{\text{facing}}$ is the solar facing area, $c_{\text{refl}}$ is the reflectivity constant, and $m_B$ is the mass of the satellite.

\subsection{Gravitational Effects from Celestial Bodies}
The presence of other celestial bodies; namely the Sun, Phobos, Deimos, and Jupiter, affect the satellite's motion in AMO through their gravitational pull. While small relative to the spherical harmonics perturbations, these disturbances can have a compounding effect, especially on the short-period oscillations of the spacecraft within the station-keeping window. The direction and magnitude of the perturbative acceleration varies due to the location of Mars relative to the Sun and Jupiter, and thus are primarily epoch dependent. While the moons of Mars are small, they are in relatively close proximity and demonstrate a non-negligible effect on the trajectory of the satellite from its desired orbital path, primarily in short-period motion. Interestingly, the areostationary orbit lies radially between the two moons, where the semi-major axis of Phobos orbiting Mars is roughly 9,400km, Deimos is 23,500km, and an AMO satellite orbits with a semi-major axis of 20,427.7km at an approximate average altitude of 17,000km.

An expression for the acceleration due to the presence of the Sun, Jupiter, Phobos, and Deimos is given by the same governing equation, with a variation on the gravitational constant and position vectors. The following defines the four expressions; one for each celestial body of concern;~\cite{Konopliv2006,Weiss2018}
\begin{equation*}
	\mbf{a}^i_a = \mu_i\left(\frac{\mbf{r}_a^{ic}}{\norm{\mbf{r}_a^{ic}}^3} - \frac{\mbf{r}_a^{iw}}{\norm{\mbf{r}_a^{iw}}^3}\right),
\end{equation*}
where $i$ is the index of the celestial body (e.g., $i = s$, where s corresponds to the gravitational center of the Sun), and $\mu_{i}$ is the gravitational parameter of the Sun, Jupiter, Phobos, and Deimos. The term $\mbf{r}_a^{ic}$ describes the position of the celestial bodies relative to the spacecraft center of mass expressed in $\mathcal{F}_a$. Likewise, the term $\mbf{r}_a^{iw}$ represents the position of the celestial bodies relative to a fixed point $w$ at the center of Mars expressed in $\mathcal{F}_a$.

\begin{figure}[t!]
	\includegraphics[width=0.8\linewidth]{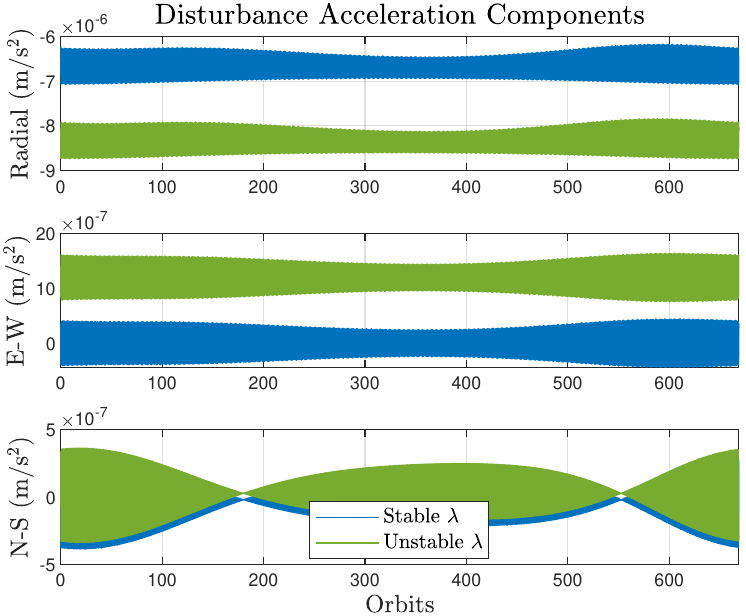} 
	\centering
	\caption{A time history of disturbance acceleration components (m/s$^2$) in Hill's Frame $\mathcal{F}_h$ at a stable longitude ($17.92^\circ$~W) and an unstable longitude ($148^\circ$~W) over 668.6 orbits, which is 687 Earth days, the length of Mars' orbital period around the sun.}\label{fig:distForce}
\end{figure}

\subsection{Total Perturbations}
All disturbances defined in this section can be added together to yield a single disturbance acceleration term acting on the spacecraft. Specifically, this is written as
\begin{align}\label{eq:totalPert}
\mbf{a}_a^p = \mbf{a}_a^{\textrm{grav}} + \mbf{a}_a^{\textrm{srp}} + \mbf{a}_a^{\textrm{sun}} + \mbf{a}_a^{\textrm{phobos}} + \mbf{a}_a^{\textrm{deimos}} + \mbf{a}_a^{\textrm{jupiter}},
\end{align}
where $\mbf{a}_a^p$ is the perturbation force per unit mass as defined in~\eqref{eq:EoM1}, and all accelerations on the right-hand side are defined in prior subsections of Section~\ref{Sec:EnvModel}. These are all resolved in $\mathcal{F}_a$ and thus can simply be added together to form the term defining all perturbations to the nominal Keplerian equations of motion. Figure~\ref{fig:distForce} shows the time history of disturbance force components over one Mars year with the satellite placed at two different longitudes; a stable one ($17.92^\circ$~W) where the along-track (E-W) spherical harmonic perturbations are zero, and an unstable one ($148^\circ$~W), where the along-track perturbations are nonzero. Note that the definition of a stable or unstable longitude is entirely dependent on the satellite's location above the planet, and other celestial perturbations will cause the satellite to drift even when placed with perfect initial conditions at a stable longitude. 

\begin{figure}[t!]
	\centering
	\begin{subfigure}[]{0.5\textwidth}
		\includegraphics[scale=0.5]{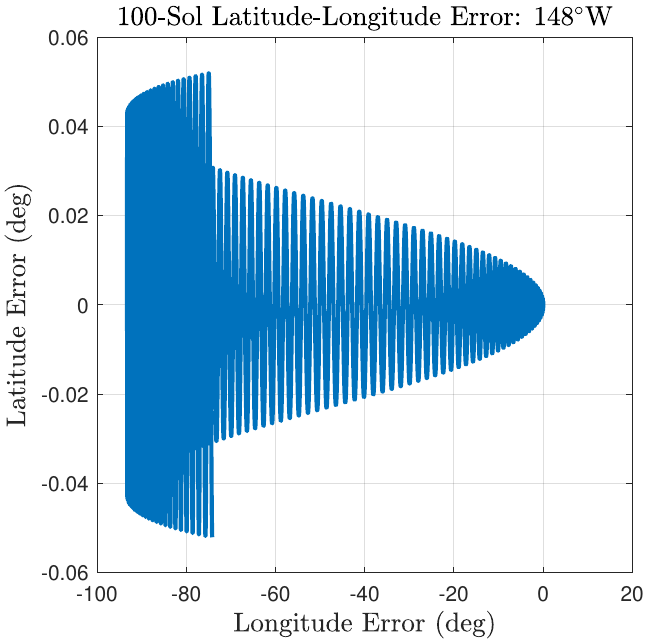} 
		\caption{}\label{fig:unstable_LatLongError}
	\end{subfigure}%
	\begin{subfigure}[]{0.5\textwidth}
		\includegraphics[scale=0.5]{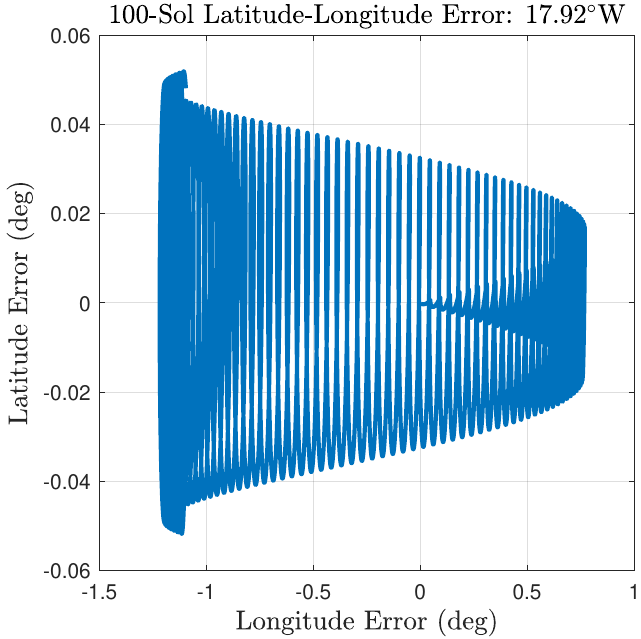} 
		\caption{}\label{fig:stable_LatLongError}
	\end{subfigure}%
	\centering
	\vspace{-4pt}
	\caption{Plots of uncontrolled error in latitude-longitude position relative to stationary orbit at (a) unstable ($148^\circ$~W) and (b) stable ($17.92^\circ$~W) longitudes over 100 orbits.}\label{fig:LatLongError}
\end{figure}

Figure~\ref{fig:LatLongError} shows the uncompensated motion of the satellite over 100 orbits subject to these perturbations at the same stable and unstable longitudes as those studied in Figure~\ref{fig:distForce}. It is worthy to note the pronounced difference in the drift of the satellites at different longitudes ($<1.5^\circ$~drift in longitude from a stable $\lambda$ vs. $>80^\circ$~drift from an unstable $\lambda$), further pointing to the longitudinal dependence of the perturbations and thus the magnitude of drift if left uncompensated. It is also clear that the satellite trajectory reaches some limit cycle through oscillations about the stable longitude in Figure~\ref{fig:stable_LatLongError}. This limit cycle is believed to be due to the non-zero perturbations in the radial ($h_1$) direction. Meanwhile, the satellite trajectory when initially placed over an unstable longitude eventually crosses through an equilibrium longitude and also begins to oscillate about that longitude. The uncompensated motion motivates the importance of assessing the proposed control method in this work at various longitudes, which in turn could be an important factor in determining future ground-based mission locations. 

\section{Predictive Control Policy}\label{sec:MPC}
Model predictive control is a tractable approach for station keeping of satellites given its ability to guarantee state and control constraints are met while exploiting knowledge of expected perturbations over a receding time horizon. This enables the controller to act before constraints may be violated, rather than simply reacting to errors as they occur. A nonlinear model predictive control (nMPC) policy is used in this application, adapted from the nMPC station-keeping policy in Ref.~\cite{Halverson2023}. 

The nMPC approach takes advantage of flexibility in selection of a cost function as opposed to the linear quadratic MPC policies implemented in Refs.~\cite{Caverly2020_IEEE,Halverson2021}, which utilize cost functions analogous to a recursive finite-horizon LQR optimization problem. Specifically, the nMPC policy solves the optimization problem given by
\begin{equation}
\label{eq:nMPCcost}
\min_{\mathcal{U}_t} \hspace{5pt} \sum_{k=0}^{N-1}\norm{\mbf{u}_{k|t}}_1, 
\end{equation}
subject to
\begin{align}
&\mbf{x}_{k+1|t} = \mbf{f}_\text{d}(\mbf{x}_{k|t},\mbf{u}_{k|t},\mbf{w}_{k|t}), \label{eq:nonlcon1} \\
&\mbf{x}_{0|t} = \mbf{x}(t), \nonumber \\
& \mbf{w}_{k|t} = \mbfhat{w}_t(t+k\Delta t), \label{eq:nonlcon2} \\
&\delta\mbf{x}_{\text{min}} \leq \delta\mbf{x}_{k|t} \leq \delta\mbf{x}_{\text{max}}, \quad 0 \leq k \leq N, \label{eq:nonlcon3}  \\
&\mbf{u}_{\text{min}} \leq \mbf{u}_{k|t} \leq \mbf{u}_{\text{max}}, \quad 0 \leq k \leq N-1,\label{eq:nonlcon4} 
\end{align}
where $N$ is the number of discrete time steps that make up the prediction horizon, $\mathcal{U}_t = \{\mbf{u}_{0|t},\ldots,\mbf{u}_{N-1|t}\}$ is the control input across the prediction horizon with $\mbf{u}_{k|t}^\trans = [ u_{1,k|t} \,\,\, u_{2,k|t} \,\,\, u_{3,k|t}]$, $\hat{\mbf{w}}_t(j)$ is the open-loop predicted disturbance column matrix at time $j$ based on data at time $t$, $\delta\mbf{x}_{k|t}$ denotes the spacecraft state error relative to its desired position and velocity in $\mathcal{F}_h$ at time step $k$, and~\eqref{eq:nonlcon1} is a discrete-time representation of the continuous-time nonlinear equations of motion derived using~\eqref{eq:EoM1}. Specifically, $\mbf{f}_\text{d}(\mbf{x}_{k|t},\mbf{u}_{k|t},\mbf{w}_{k|t})$ is defined via a zero-order hold discretization of the continuous-time state space model given by
\begin{equation*}
    \dot{\mbf{x}} = 
    \mbf{f}(\mbf{x}(t),\mbf{u}(t),\mbf{w}(t)) = 
    \begin{bmatrix}
    \dot{\mbf{r}}^{cw}_a\\
    \ddot{\mbf{r}}^{cw}_a\\
    \end{bmatrix} = 
    \begin{bmatrix}
    \frac{d}{dt}\mbf{r}_a^{cw}\\
    -\mu \frac{\mbf{r}_a^{cw}}{\norm{\mbf{r}_a^{cw}}^3} + \mbf{w}(t) + \frac{1}{m_B}\mbf{u}(t)\\
    \end{bmatrix},
\end{equation*}
where the state vector $\mbf{x}\in\mathbb{R}^6$ consists of the position and velocity vectors of the spacecraft, $\mbf{r}_a^{cw}$ and $\dot{\mbf{r}}_a^{cw}$, respectively, and $\mbf{w}(t)$ comes from the total perturbations defined by~\eqref{eq:totalPert}. The solution to the optimization problem in~\eqref{eq:nMPCcost} is given by $\mathcal{U}_t^* = \{\mbf{u}_{0|t}^*,\ldots,\mbf{u}_{N-1|t}^*\}$ and the control input applied at time $t$ is chosen as $\mbf{u^*} = \mbf{u}_{0|t}^*$. The control input is applied as a zero-order hold, such that the satellite would provide the desired thrust, $\mbf{u^*}$, continuously over the entire time step.

The instantaneous saturation limit on thrust is defined as $\mbf{u}_\text{min}\leq\mbf{u}\leq\mbf{u}_\text{max}$, where $\mbf{u}_\text{min} = -\mbf{u}_\text{max}$ and $\mbf{u}_\text{max}^\trans = f_\text{max}\left[1~1~1\right]$. The station-keeping window constraint is characterized by $\delta\mbf{r}_\text{min}\leq\delta\mbf{r}_h\leq\delta\mbf{r}_\text{max}$, where $\delta\mbf{r}_\text{max}^\trans = \left[\infty ~\Bar{r}\tan(\lambda_\text{max})~\Bar{r}\tan(\phi_\text{max})\right]$, $\delta\mbf{r}_\text{min} = -\delta\mbf{r}_\text{max}$, and $\Bar{r} = \norm{\Bar{\mbf{r}}_a^{cw}}$ is the nominal AMO semi-major axis. This station-keeping window constraint is incorporated with the entire state constraint such that $\delta\mbf{x}_\text{max} = \left[\delta\mbf{r}_\text{max}^\trans~\delta\mbf{v}_\text{max}^\trans\right]^\trans$, where $\delta\mbf{v}_\text{max}^\trans = \left[\infty~\infty~\infty\right]$.   

In essence, this control policy aims to find the minimal control inputs over a receding-time prediction horizon, $N$, computed at time step $k$ subject to specific constraints. The expected state response is propagated forward to forecast the system behavior through the nonlinear dynamics in~\eqref{eq:nonlcon1}. The dynamics are subject to the expected perturbations in~\eqref{eq:nonlcon2} as well as the control input solved for by the optimization problem. The nMPC policy selects the control inputs that minimize the cost function given the expected future response while ensuring the state and control constraints in~\eqref{eq:nonlcon3} and~\eqref{eq:nonlcon4} are satisfied. Specifically, the satellite is constrained to be bounded within $\delta \mbf{x}_\text{min}$ and $\delta \mbf{x}_\text{max}$ while using control inputs in the range of $\mbf{u}_\text{min}$ to $\mbf{u}_\text{max}$. The control input at the current time step, $k$, is then selected and applied to the satellite system. Note that the MPC policy minimizes \textit{all} control inputs over the entire prediction horizon, however only the first input is applied to the system. The optimization is repeated at the next time step given the actual state measurement, and the optimization is completely re-solved with this measurement as the initial condition for the prediction and calculation of expected disturbances. A complete description of variations of predictive control and its application can be found in Ref.~\cite{rawlings2017model}.

The optimization problem in~\eqref{eq:nMPCcost} directly minimizes the $\Delta v$ applied to the spacecraft over a finite prediction horizon through minimizing the $\mathcal{L}^1$ norm of the control input vector. The objective function minimizes the sum of the absolute value of thrust input at each time step over the prediction horizon where $\mathcal{L}^1(\mbf{u}_{k|t}) = \norm{\mbf{u}_{k|t}}_1 = \abs{u_{1,k|t}} + \abs{u_{2,k|t}} + \abs{u_{3,k|t}}$. This is equivalent to $(m_B\Delta v)/\Delta t$ in this case given the assumption that six bi-directional thrusters are mounted on the spacecraft in three mutually orthogonal directions in $\mathcal{F}_h$. Specifically, the total $\Delta v$ is calculated as
\begin{equation*}
    \Delta v_\text{tot} = \frac{1}{m_B}\sum_{i=0}^{M}\mathcal{L}^1(\mbf{u}_{i|t})\Delta t,
\end{equation*}
where $\Delta t$ is the discretization time step length, $i$ denotes each time step in the simulation, and $M$ is the total number of time steps in the entire simulation.

\subsection{nMPC Solve Time \& Implementation}
Real-world implementation is an important topic to discuss, as spacecraft computational power is often limited and many uncertainties exist, regardless of fidelity of the simulation environment the control policy is tested on. The policy in~\eqref{eq:nMPCcost} is implemented in a station-keeping simulation using MATLAB's \texttt{fmincon} solver~\cite{fmincon}. This simulation is performed on a laptop equipped with an Intel Core i7-12700H processor to provide a baseline for the computational time required to solve the policy in~\eqref{eq:nMPCcost}. Performing a 25 timestep simulation with a prediction horizon length of $N = 14$ and a timestep length of $\Delta t = 1$~hour, the solver took an average of $4.287$~seconds with a standard deviation of $0.166$~seconds. This time-to-solution equates to approximately $0.12\%$ of the timestep length, which illustrates that the MPC policy has significant time to find a solution for the control input before the next timestep. 

It should be noted there are several tradeoffs with regards to computation time versus performance. Two distinct examples include:
\begin{enumerate}
    \item Prediction horizon length. As the prediction horizon, $N$, increases, it is expected that performance with regards to the cost function will improve, as more information is available to the control policy (dependent on the accuracy of the prediction model). However, the computation time increases significantly with each increase in the prediction horizon length. Fortunately, the results presented in this paper do not rely on long prediction horizons and can be solved in a fraction of the length of the discrete timestep.
    \item MPC policy type and formulation. The nonlinear MPC policy in this work includes an accurate nonlinear prediction model and a cost function that directly minimizes $\Delta v$. Alternatively, linear MPC implementations allow for the solution to be found efficiently via quadratic programming in a convex optimization approach, but result in greater $\Delta v$ usage. This tradeoff is described extensively in Ref.~\cite{Halverson2023}, and methods for solving MPC policies using linear models can be found in Ref.~\cite{Muske1993}.
\end{enumerate}
Further, this control policy could be massively improved with regards to solve time with coding practices used for space vehicle computation, such as coding in C with a more efficient optimization framework. It is expected that the speed-up in computing time through this would be offset by the limited computational power available to on-board radiation-hardened spacecraft processing units. Therefore, the computation time described in this section should be interpreted more as a rough proof of feasibility, rather than a rigorous assessment of the on-board computational resources required.

Finally, there may be practical concerns regarding the solvability of the MPC policy, where theoretical guarantees regarding finding a solution in finite time---especially in the nMPC policy---may not exist. This problem may be approached in several different ways, such as including soft constraints to allow for a solution in cases where unanticipated events cause the problem to become infeasible~\cite{rawlings2017model}, the use of an auxiliary stabilizing regulation control policy (e.g., LQR, PD, etc.) to supplement the MPC policy if a solution is not found in time, using the previous timestep's solution at the appropriate position in the prediction horizon, or formulating an MPC policy that is robust to early termination itself~\cite{Hosseinzadeh2023}. At least one of these approaches would be implemented on-board a spacecraft implementing the proposed nMPC station-keeping policy. 

\section{Station Keeping $\Delta v$ vs. Longitude}\label{sec:longSweep}
One of the earliest works that investigated the expected $\Delta v$ required for station keeping of AMO satellites primarily characterized the longitude evolution of a satellite and introduced a simple method for controlling the satellite's mean longitude~\cite{Silva2013}. Specifically, Ref.~\cite{Silva2013} quantified the expected $\Delta v$ for longitudinal control of an AMO satellite using the equation
\begin{equation}\label{eq:est_dv}
    \Delta v = \frac{V_s \cdot \Delta D}{3},
\end{equation}
where $\Delta D$ is the expected annual dimensionless longitudinal drift rate change and $V_s$ is the nominal areostationary velocity. This equation provides the yearly $\Delta v$ required to perform solely tangential longitudinal evolution station-keeping maneuvers for control of the mean longitude, where the satellite is subjected to gravitational and solar radiation pressure perturbations. The reader is encouraged to see Ref.~\cite{Silva2013} for a complete description of how optimal longitudes were determined and the calculations of expected $\Delta v$ required for AMO satellites due to these perturbations.

\begin{figure}[t!]
	\includegraphics[scale=0.75]{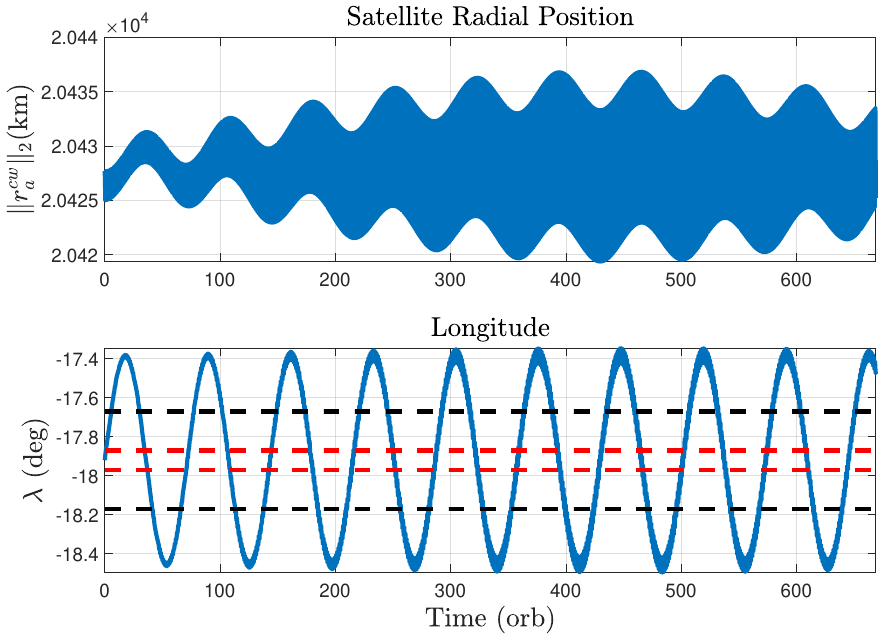} 
    \hspace{-15pt}
	\caption{Satellite response to a na\"ive station-keeping policy where along-track and cross-track perturbations are canceled out leaving only radial perturbations. Image shows satellite radial position and longitude (blue) vs number of orbits, and longitudinal bounds of an imaginary station-keeping window at $\pm 0.05^\circ$ (red) and $\pm 0.25^\circ$ (black).}\label{fig:radial_ignored}
\end{figure}

Note that in the use of these perturbations, Ref.~\cite{Silva2013} only considered the East-West (i.e., along-track or tangential) perturbation acceleration and its effect on the satellite motion.  For proper longitudinal station keeping, radial perturbations (and other environmental effects) must also be taken into account. See Figure~\ref{fig:radial_ignored} for an example of improper station keeping at a stable longitude where radial perturbations are ignored in the crude station-keeping policy, causing the satellite to drift too far outside of the  prescribed station-keeping windows.  In this example, the along-track and cross-track perturbations are directly canceled, leaving only the drift due to radial disturbances. A larger station-keeping window might contain the cyclic response of the satellite longitudinal motion for a certain period of time, however this would come at a cost with regards to ground communications, where the satellite may not be able to be approximated as fixed in the sky due to $>1^\circ$ drift. This constraint due to ground communications is further discussed in Section~\ref{sec:SMP}.

\begin{figure}[t!]
	\includegraphics[width=0.95\linewidth]{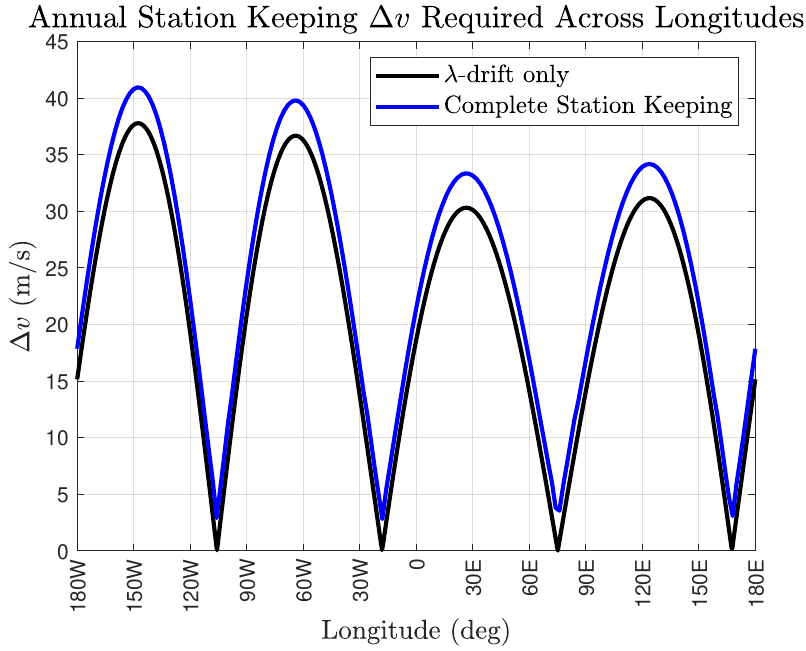} 
	\centering
	\caption{Annual (355-orbit) $\Delta v$ required for areostationary station keeping across all longitudes. The results in black ($\lambda$-drift only) use the station-keeping policy in Ref.~\cite{Silva2013}. The results in blue (complete station keeping) use the nonlinear MPC policy introduced in Section~\ref{sec:MPC}.}\label{fig:longSweep_combined}
\end{figure}

Ref.~\cite{Silva2013} also introduced---to the best of the authors' knowledge---the first complete expected values of annual (Earth-year) $\Delta v$ required for longitudinal station keeping in AMO using the equation in~\eqref{eq:est_dv}. Namely, they defined the locality of equilibrium points using the higher-order spherical harmonics, as described in Section~\ref{sec:spherHarm}, to define two stable ($17.92^\circ$W and $167.83^\circ$E) and two unstable ($105.55^\circ$W and $75.34^\circ$E) longitudinal equilibrium points. A re-creation of the longitude sweep data for this $\Delta v$ estimate via quantification of longitudinal drift only is shown in Figure~\ref{fig:longSweep_combined}. 

The following results utilize the dynamics described in Section~\ref{Sec:EnvModel} to quantify the annual $\Delta v$ required for station keeping of an areostationary satellite. We use the station-keeping policy introduced in Section~\ref{sec:MPC}, which accounts for complete three-degree-of-freedom control including control of longitudinal and latitudinal evolution within a prescribed station-keeping window. 

\subsection{Simulation and Controller Parameters}
For a sweep of the expected annual $\Delta v$ required for station keeping of a satellite in AMO at various longitudes, a constant set of controller and simulation parameters are selected. Each simulation is performed for 615 orbits such that the satellite's transient trajectory to the North-South boundary of the station-keeping window is removed. On average, this orbital inclination evolution takes 260 orbits. Thus, 355 orbits of $\Delta v$ accumulation data is collected, which corresponds to one Earth-year worth of time. 

To define the solar radiation pressure perturbation, the parameters $C_\text{srp} = 4.5\times10^{-6}$~N/m$^2$, $S_\text{facing} = 37.5$~m$^2$, $c_\text{refl} = 0.6$, and $m_B = 3940$~kg are selected. These parameters are chosen to be similar to that of a GEO satellite. The full simulation begins at an epoch of January 1st, 2000 at noon GMT.  The limit $f_\text{max} = 50$mN was chosen to be within the bounds of allowable thrust for an electric propulsion system. The station-keeping window is defined with limits in latitudinal and longitudinal drift of $\phi_\text{max} = \lambda_\text{max} = 0.05^\circ$, respectively, or a maximum translational drift of $17.83$~km in latitude and longitude. The limit of a square station-keeping window of $0.05^\circ$ is chosen to be similar to that of existing GEO missions, and limit the maximum drift such that ground-based sensors wouldn't require much slewing (or any at all). Finally, the prediction horizon was chosen as $N = 23$~hours to provide a meaningful look into the future while taking into account diminishing returns on performance at the expense of increased computation time. 

\subsection{Sweep Results}
Results showing the annual $\Delta v$ required for station keeping of an areostationary satellite using the MPC policy in Section~\ref{sec:MPC} and the simulation parameters defined above can be seen in Figure~\ref{fig:longSweep_combined}. The $\Delta v$ required at each longitude is slightly greater than that estimated in Ref.~\cite{Silva2013}. This is largely due to the fact that the estimated $\Delta v$ only considers the longitudinal drift that arises from the along-track perturbation due to non-spherical gravitational perturbation. The nMPC station-keeping policy is able to utilize drift within the station-keeping window defined by constraints on latitude and longitude in order to minimize the required fuel, subject to the relevant perturbations in the AMO environment. 

\begin{figure}[t!]
	\includegraphics[width=0.95\linewidth]{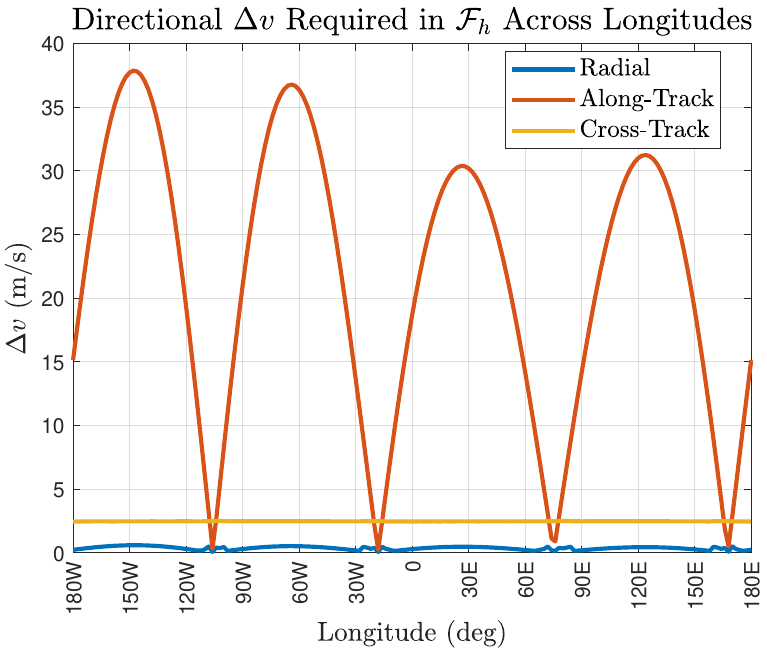} 
	\centering
	\caption{Annual $\Delta v$ required for areostationary station keeping across all longitudes using a nonlinear MPC policy, with $\Delta v$ split into directional components in Hill's frame.}\label{fig:longSweep_split}
\end{figure}

The same results using the MPC policy across all longitudes is shown in Figure~\ref{fig:longSweep_split}, where $\Delta v$ requirements are split into radial, along-track, and cross track components. Ultimately, the cross-track (North-South) $\Delta v$ is constant at any longitude, given the inclination evolution is largely driven by perturbations that are not longitudinally dependent (i.e. other celestial bodies and solar radiation pressure). Furthermore, the accumulated radial $\Delta v$ is small relative to the other directions due to the general inefficiency in thrusting radially to change the along-track motion. This phenomena is generally understood in orbital mechanics, however it is clear that through MPC's optimization problem---given the capability to thrust in the radial direction---there is some fuel cost savings in thrusting radially, especially at non-equilibrium longitudes. Moreover, the along-track and radial thrust do not simply cancel out the perturbations in those directions. As seen in Figure~\ref{fig:distForce}, the radial perturbations are actually greater than (sometimes by orders of magnitude) the along-track perturbations. This indicates that the along-track thrust not only reacts to perturbations in that direction, but also to those in the radial direction that affect the satellite's along-track motion.

\section{Example Application: Southern Meridiani Planum}\label{sec:SMP}
The control policy outlined in Section~\ref{sec:MPC} is implemented in an example AMO station-keeping scenario, where a mission designer may be interested in placing a satellite over a site of interest. The southern Meridiani Planum is suggested as a candidate for a crewed mission to the surface of Mars~\cite{clarke2017southern}. The landing site is one of the most well-documented regions of Mars that boasts potential water resources and diverse science features while being cited as an accessible and safe candidate landing location. The proposed landing site at the southern Meridiani Planum is precisely located at $5^\circ 30' 2.46''$W and $3^\circ 5' 4.84''$S, making it a great location for placement of an equatorial areostationary satellite overhead. This section explores the design considerations, controller tuning process, and some results for station keeping of a satellite placed in AMO at a longitude $\lambda = 5.5^\circ$W.

\subsection{Design Considerations}
There are several considerations from a mission design perspective one would need to take into account for placement of an areostationary satellite that supports a crewed mission to the surface of Mars. First and foremost, elongating satellite lifetime is of one of the most important objectives. This is primarily influenced by the fuel required to maintain the satellite over its desired longitude with minimal inclination evolution. As such, a station-keeping window is prescribed where the satellite is able to drift off from its desired position directly overhead in order to minimize fuel, while being small enough to ensure the communication capabilities are not diminished. For simplification of mission design, the size of the station-keeping window should be such that the ground-based antenna would not need to move in order to fully take advantage of the stationary orbit. It is well understood that there is a direct trade-off between directional antenna gain and beam-width~\cite{harrington1958gain}, and as such, the smaller a station-keeping window, the higher the gain a ground-based antenna beam would be able to provide. This allows for greater data transfer and telecommunication rates. 

From the satellite perspective, the length of the prediction horizon in the MPC policy is often the greatest driver of station keeping fuel requirements. It is generally expected that with an increase of the prediction horizon, the better the performance of the controller given that it can `see' further into the future. This isn't without diminishing returns, however. The computation time for the MPC optimization required increases exponentially with an increase in prediction horizon length. Moreover, the longer the prediction horizon, the more uncertainty is to be accumulated due to an imperfect model such that a longer prediction time horizon is expected to be less accurate. Thus, a prediction horizon should be chosen to be long enough to provide the MPC policy sufficient information for the expected future motion of the satellite while also being short enough so the computation time does not exceed the time the satellite has to solve for a control input. 

In this work, the control discretization time-step is one hour in a zero-order hold, i.e., a new control input is selected each hour of the mission and held constant over that hour. Generally, it can be expected---with some optimization for on-board processing---that the control system computer can easily solve for the control input within an hour. However, this long discretization time-step provides a potential challenge where the optimization problem is solved using `old data', such that the control input is selected up to an hour before it is actually implemented, for example. This sort of latency would need to be considered in the positional determination and control system design specific to the satellite mission. 

Other design considerations include thruster placements, gimbaling capabilities, possible quantization to minimize on-off pulses and increase thruster lifetime, and coupled attitude control. These control challenges are explored in the GEO station keeping case in great detail in Ref.~\cite{Caverly2020_IEEE}. There are other considerations one may take into account regarding power draw from thruster implementation, such that the satellite remains power positive. The MPC formulation could be modified to include a function for minimizing power draw from thrusters in the cost function. Further, this work assumes the satellite is placed in a perfect areostationary orbit from the very beginning of the simulation. Ultimately, proper placement in a circular orbit at the correct longitude may take significant fuel usage after the satellite reaches Mars. 

\subsection{Tuning MPC Parameters}
\begin{figure}[t!]
	\includegraphics[width=0.95\linewidth]{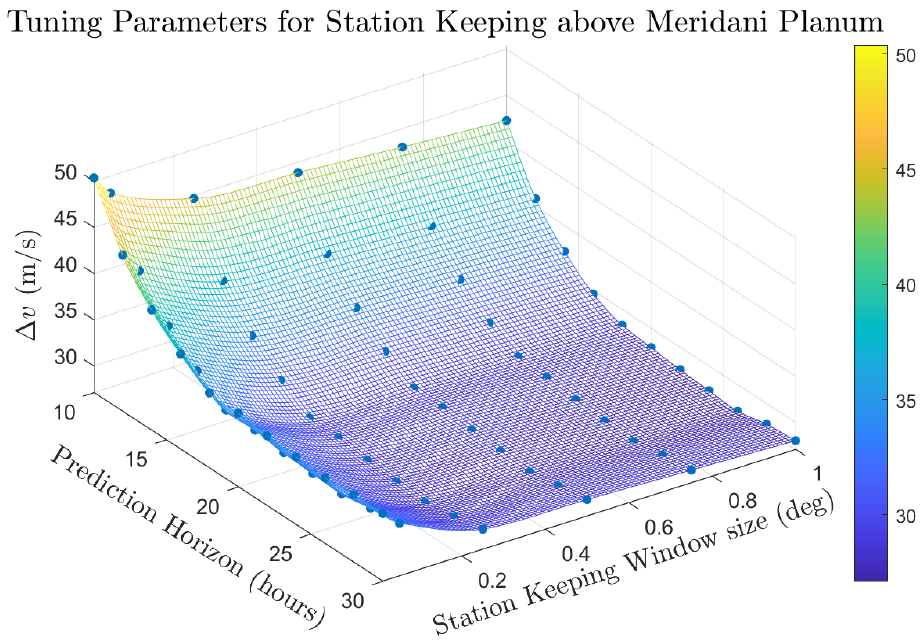} 
	\centering
	\caption{Station-keeping $\Delta v$ required as a function of prediction horizon length and station-keeping window size with a nonlinear MPC policy applied in areostationary orbit above Southern Meridiani Planum. Simulations are performed for 668.6 orbits, or the length of one Martian year. A cubic meshgrid is used to interpolate between data points and approximate the surface shown.}\label{fig:SMP_tuning}
\end{figure}

Results from tuning the MPC policy for a satellite in AMO directly above the Southern Meridiani Planum can be seen in Figure~\ref{fig:SMP_tuning}. Simulations were performed for one Martian year, or 668.6 orbits in AMO, each starting at an epoch of January 1st, 2040 at noon GMT. This starting date was chosen for this particular study given the potential for a real-world implementation. The satellite parameters chosen are the same as those in Section~\ref{sec:longSweep}. The MPC policy has a discretization length $\Delta t = 1$~hour, and a thrust force saturation limit of $f_\text{max} = 50$~mN. Two important MPC tuning parameters are assessed through sweeps, including  the prediction horizon length and station-keeping window size. It is expected for the $\Delta v$ to trend downwards as both the prediction horizon and station-keeping window size increase, since the MPC policy has more future information, and there is more space available for the satellite to drift within the station-keeping window.

Figure~\ref{fig:SMP_tuning} shows the accumulated $\Delta v$ over one Mars year versus variations in the MPC prediction horizon between 10 and 30 time-steps, and a station-keeping window size between $\pm 0.01^\circ$ and $\pm 1^\circ$. Note that the station-keeping window is defined as a square deviation from the desired longitude and $0^\circ$ latitude. The $\Delta v$ required decreases as the prediction horizon and station-keeping window size both generally increase, as expected. For each case, however, there are diminishing returns. As discussed earlier, an increase in prediction horizon length directly causes a significant increase in the computation time for the optimization problem. Further, the station-keeping window being too large could be cause for the satellite to fall out of a fixed ground station's beam-width. Taking these limitations into account, a prediction horizon of $N = 26$ and a station-keeping window size of $\delta\phi_\text{max} = \delta\lambda_\text{max} = 0.25^\circ$ are chosen as the best MPC parameters for this given longitude and mission. 

A similar tuning sweep could be completed for an AMO satellite at any other longitude to optimize the MPC policy for the specific environment. Further design choices regarding uncertainty of the prediction model and robustness to un-modeled disturbances should be included in any real-world mission design. Generally, the annual $\Delta v$ requirements will follow that of the results shown in the longitude sweep in Figure~\ref{fig:longSweep_combined}. In this specific case, the expected $\Delta v$ required for station keeping over 355 orbits---one Earth year---at a longitude of $5.5^\circ$W was $16.303$ m/s with the tuning parameters utilized in the longitude sweep ($N = 23$ and $\delta\lambda_\text{max} = \delta\phi_\text{max} = 0.05^\circ$). Given the new controller parameters specified from this tuning sweep, the $\Delta v$ required for 355 orbits above the Southern Meridiani Planum is $14.987$m/s, a direct fuel savings of $\approx 10\%$. It is expected, however, that the $\Delta v$ requirement will increase as the satellite reaches the north-south boundary of the station-keeping window due to the inclination evolution. 

\begin{figure}[t!]
	\begin{subfigure}[]{1\textwidth}
        \centering
		\includegraphics[scale=0.65]{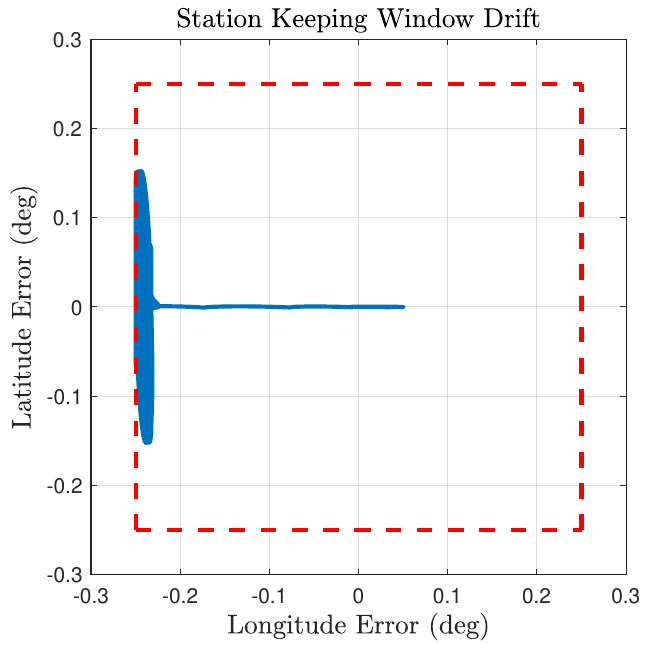} 
		\caption{}\label{fig:SMP_Window_over}
	\end{subfigure} \\
	\begin{subfigure}[]{1\textwidth}
        \centering
		\includegraphics[scale=0.65]{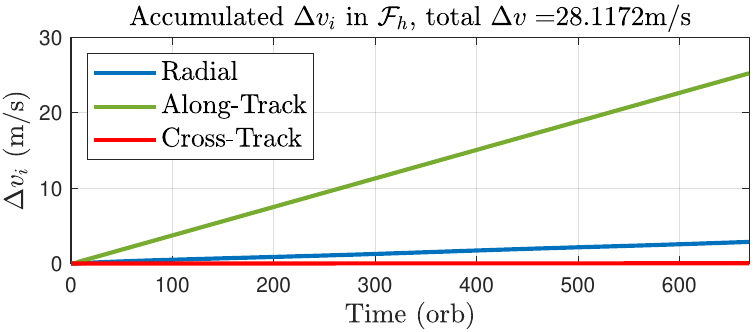} 
		\caption{}\label{fig:SMP_DV_over}
	\end{subfigure} \\
    \begin{subfigure}[]{1\textwidth}
        \centering
		\includegraphics[scale=0.65]{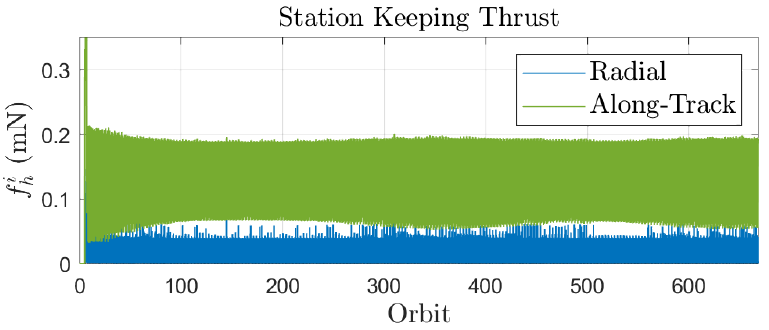} 
		\caption{}\label{fig:SMP_thrust_over}
	\end{subfigure}%
	\caption{Station-keeping results over Southern Meridiani Planum, with (a) the satellite trajectory (blue) within the station-keeping window (red), (b) accumulated $\Delta v$ over one Mars year or 668.6 orbits, and (c) a time history of radial and along-track thrust required. A prediction horizon of 26 hours is used in the MPC policy.}\label{fig:SMP_Expanded}
\end{figure}

A simulation of an AMO satellite placed directly over the Southern Meridiani Planum is performed for one Mars year, or 668.6 orbits, with the same MPC controller parameters defined above ($N=26$~hours and $\delta\lambda_\text{max} = \delta\phi_\text{max} = 0.25^\circ$). Results from this tuned control policy are shown in Figure~\ref{fig:SMP_Expanded}. The $\Delta v$ required for station keeping at this specific longitude of $5.5^\circ$~W is $28.1172$~m/s (mission length is roughly double that which was used for tuning the MPC policy). Given that the satellite is located at a non-equilibrium longitude, the amount of drift the satellite can use within the station-keeping window is minimal, as it is essentially being `dragged' towards the stable equilibrium longitude at $17.98^\circ$W. This trajectory can clearly be seen in Figure~\ref{fig:SMP_Window_over}, which shows the complete history of the latitude and longitude error over the satellite lifetime of one Mars year. Further, there is no thrust required in the cross-track direction for the duration of this simulation, as the satellite has yet to reach the North or South boundaries of the station-keeping window. For the same reason, the $\Delta v$ accumulated in the cross-track direction is also zero. 

It is interesting to note that the thrust required in the along-track direction is always non-zero. Regardless, the maximum thrust magnitude does not generally exceed 0.2~mN with few exceptions. This could make way for a slightly more complex control policy with limits on thruster cycles and quantization, which would call for fewer thrusts and limits to on-off cycles such that the satellite does not need to continually thrust, and the lifetime of the propulsion system itself may be extended. 

\subsection{Changes in Epoch}
\begin{figure}[t!]
	\centering
	\begin{subfigure}[]{0.9\textwidth}
		\includegraphics[scale=0.65]{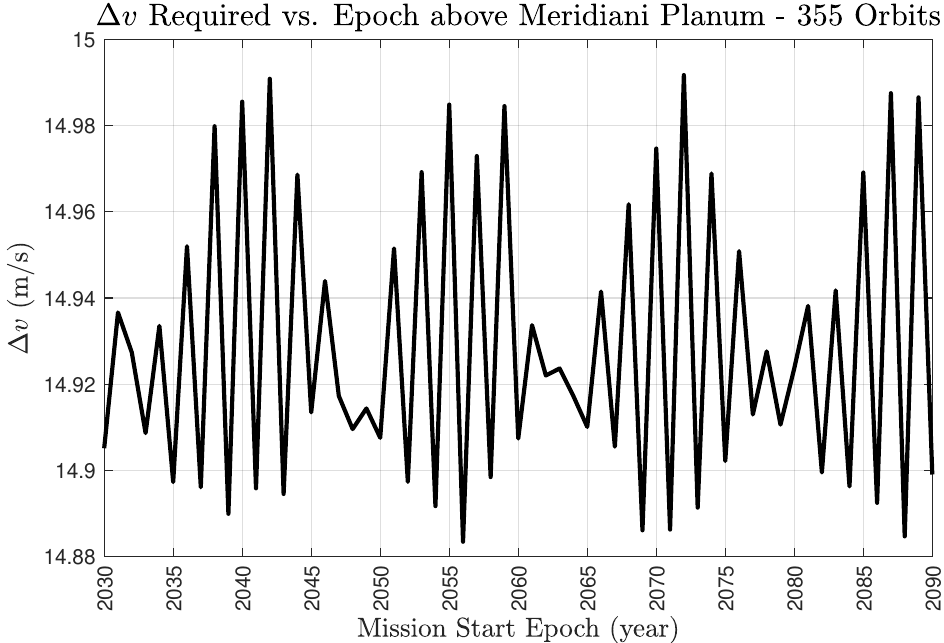} 
		\caption{}\label{fig:epoch_SMP_355}
	\end{subfigure} \\
	\begin{subfigure}[]{0.9\textwidth}
		\includegraphics[scale=0.65]{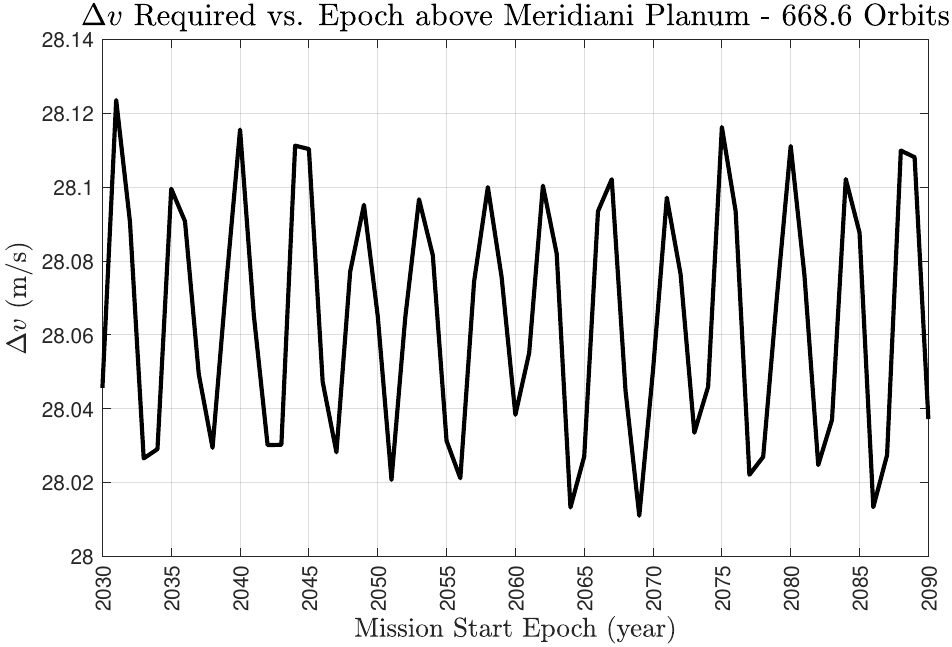} 
		\caption{}\label{fig:epoch_SMP_full}
	\end{subfigure}
	\centering
	\caption{Station-keeping $\Delta v$ required vs. mission start date on January 1st at 12:00PM UTC at the year specified. Simulations were ran for one Earth year or 355 orbits (a) and one Mars year or 668.6 orbits (b).}\label{fig:epoch_SMP}
\end{figure}

Figure~\ref{fig:epoch_SMP} shows sweeps of $\Delta v$ accumulation for varying mission start epochs between the years 2030 and 2090, each beginning on January 1st at 12:00pm UTC. Two cases are presented, where the simulation length is varying from one Earth year (355 orbits in AMO), or one Mars year (668.6 orbits). It is immediately clear that change in mission start time does not have a significant effect on the $\Delta v$ accumulated throughout the mission; regardless of its length. The difference between the greatest and smallest $\Delta v$ for each of the two simulation lengths presented is approximately $0.11$m/s. Assuming this trend continues, the potential increase in $\Delta v$ required for a worst-case mission start time compared to a best-case scenario for a 10-Earth-year-mission would be approximately 0.07\%, which is insignificant in the context of mission design and minimization of fuel requirements. 

It is interesting to note---yet unsurprising---that as mission length increases, some of the periodic nature of the $\Delta v$ changes is lost simply because more time is captured on orbit. As the mission length increases further, it is expected that epoch start time would have an equal or smaller effect on $\Delta v$ required for station keeping than the mission lengths presented here. Thus, for any AMO satellite mission placed over this specific longitude, mission start epoch has a minimal effect on the $\Delta v$ requirements for station keeping. This is due to the perturbation from non-spherical harmonics as the largest contributor to satellite drift at non-equilibrium longitudes being independent of epoch. 

\section{Alternate Design: Satellite Placement over Stable Longitude}\label{sec:stableLong}
This section explores an alternative design for an AMO satellite that supports ground operations and telecommunications at the Southern Meridiani Planum site with a satellite placed over the nearest stable longitude, rather than directly overhead. We describe considerations one may take for the satellite elevation angle from the horizon when placed at this longitude, and again explore controller tuning and results at various epochs. Ultimately this section explores the massive fuel savings that may materialize with careful selection of an AMO satellite's longitude placement.  

\subsection{Satellite Elevation from Southern Meridiani Planum}

Stationary satellites in general are particularly useful in monitoring large areas on the surface of the planet. This idea has been sufficiently explored in Ref.~\cite{Montabone2021} where the authors explicitly  state that if the orbit is maintained via station keeping--for example, via the policy introduced in Section~\ref{sec:MPC}---then it is possible to usefully observe the planet in a nearly $60^\circ$ emergence angle. Stationary satellites are also obviously useful to monitor large ground areas continuously. Ref.~\cite{Montabone2021} goes further to describe potential for quasi-global coverage via an areostationary constellation, where monitoring large ground areas excluding polar regions can be achieved by a constellation of three equatorial satellites. Four may be used for more substantial ground coverage from nearly $80^\circ$S to nearly $80^\circ$N. 

Great detail regarding specifics for design of an AMO relay satellite is found in Ref.~\cite{Edwards2016}, where a discussion on the areostationary orbiter's elevation from the surface is briefly described. Specifically, Ref.~\cite{Edwards2016} states that for any sites on the surface between $\pm 30^\circ$ of the equator and the AMO satellite's longitude matching the land site's longitude, the elevation angle is always greater than $50^\circ$. Meanwhile, Ref.~\cite{Colella2017} explores the optimal placement of three areostationary satellites over near-equilibrium points and defines elevation angle evolution without station keeping, while also exploring the ground coverage achievable by three satellites. 

\begin{figure}[t!]
	\includegraphics[width=0.5\linewidth]{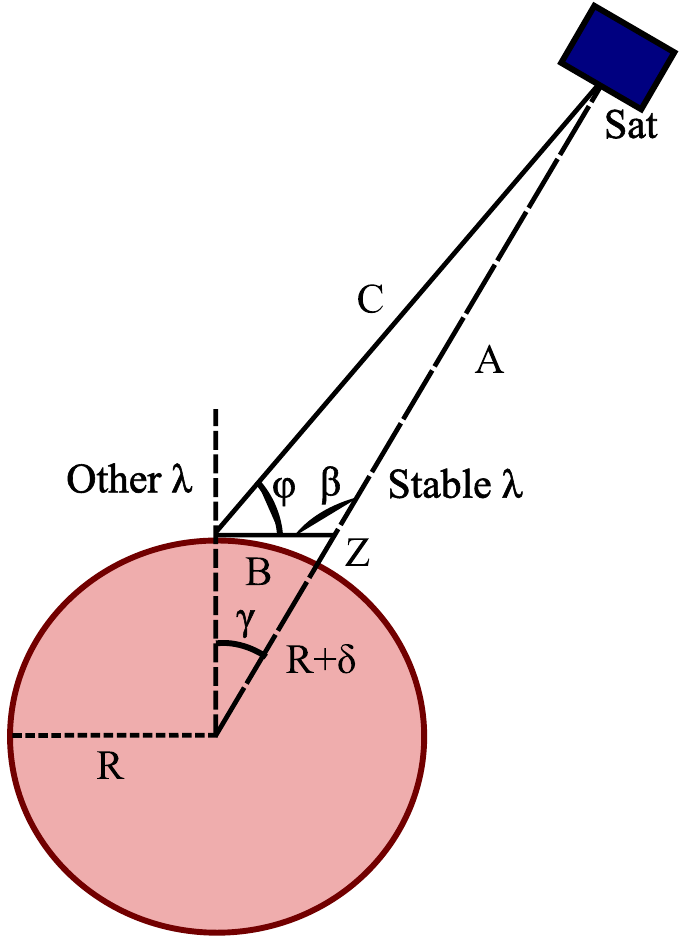} 
	\centering
	\caption{Illustration of a satellite placed over a stable longitude of Mars, where $\gamma$ is the difference in longitude between a stable $\lambda$ and some other unstable $\lambda$. The elevation angle of the satellite from the horizon at the other specific longitude is depicted as $\phi$. All other variables are useful in the calculation of $\phi$. Image not to scale.}\label{fig:elevation_work}
\end{figure}

While this study is not particularly concerned with the ground-coverage availability of equatorial areostationary satellites, we explore the elevation angle of a satellite in AMO placed directly at a stable longitude, which could support a crewed mission located on the surface at the Southern Meridiani Planum. An in-depth illustration of this configuration can be seen in Figure~\ref{fig:elevation_work}. Ultimately, the geometry problem is solved assuming only the knowledge of the difference between the stable longitude and some other longitude ($\gamma$), the radius of Mars, and the nominal altitude of an areostationary satellite. Assuming a two-dimensional problem with the satellite in an equatorial orbit and ground station located at the equator, the satellite elevation angle from the horizon ($\phi$) is calculated as
\begin{equation}
    \phi = \sin^{-1}{\left(A\frac{\sin{(\beta)}}{C}\right)},
\end{equation}
where A is the distance between the satellite and an imaginary point $Z$ at the stable longitude and along the line tangential to the horizon at the other longitude, $\beta$ is an angle to describe the triangle depicted in Figure~\ref{fig:elevation_work}, and C can be solved via law of cosines with knowledge of lengths $A$, $B$, and angle $\beta$. All intermediate values can be solved for trivially. Via these calculations, we find that the elevation angle of an AMO satellite placed at the stable longitude $\lambda_\text{stable} = 17.98^\circ$W is $\phi = 73.4^\circ$ from the horizon at the equator at the longitude of the Southern Meridiani Planum; $\lambda_\text{SMP} = 5.5^\circ$W. This indicates that the satellite is sufficiently high in the sky when placed at a stable longitude, which could allow for significant fuel savings when compared to placing the satellite directly overhead.  

\subsection{Tuning MPC Parameters}
\begin{figure}[t!]
	\centering
	\begin{subfigure}[]{0.85\textwidth}
		\includegraphics[scale=0.6]{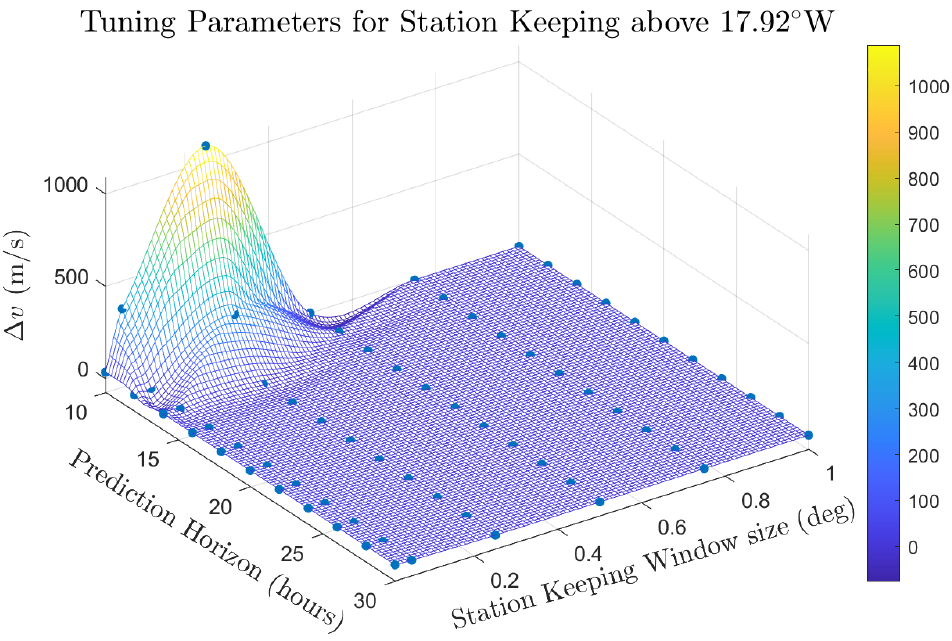} 
		\caption{}\label{fig:Stable_sweep_full}
	\end{subfigure} \\
	\begin{subfigure}[]{0.85\textwidth}
		\includegraphics[scale=0.6]{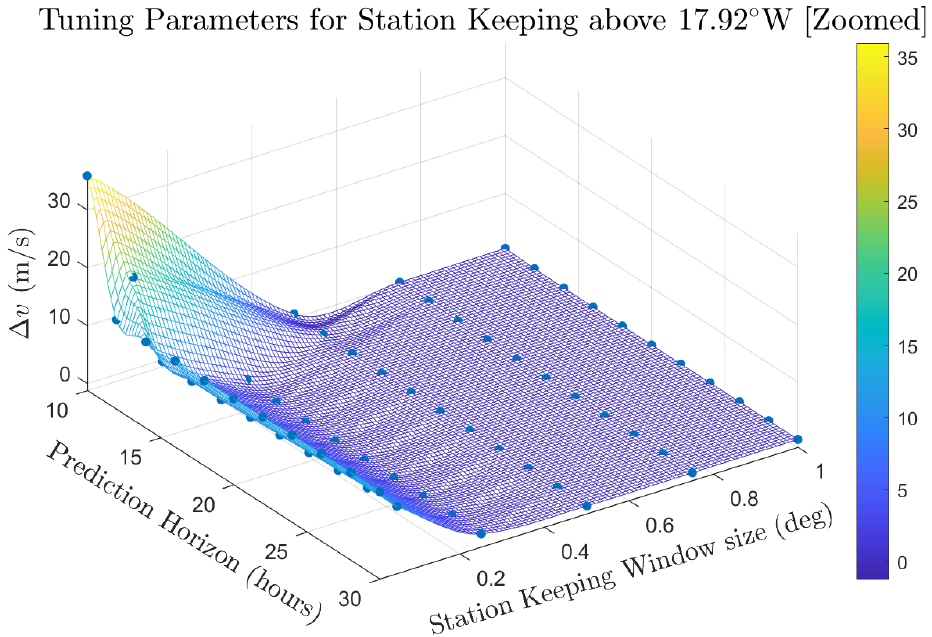} 
		\caption{}\label{fig:Stable_sweep_partial}
	\end{subfigure}
	\centering
	\caption{Station-keeping $\Delta v$ required as a function of prediction horizon length and station-keeping window size using a nMPC policy for an areostationary satellite at the stable longitude of $17.92^\circ$~W. Simulations were performed for 668.6 orbits. Sub-figure (a) shows the full suite of results, while sub-figure (b) omits data points with a $\Delta v$ greater than 50m/s to show results more clearly. A cubic meshgrid is used to interpolate between data points and provide the surface shown.}\label{fig:stable_tuning}
\end{figure}

Given that placing a satellite over a stable longitude rather than directly overhead the Southern Meridiani Planum is feasible from an elevation angle and communication ability standpoint, we explore some further design considerations here. To begin, the same tuning process used for placing the satellite directly overhead is followed, where 668.6 orbits are simulated while varying the prediction horizon length and station-keeping window size. Results from this tuning sweep are presented in Figure~\ref{fig:stable_tuning}. 

It is immediately clear that a small station-keeping window size ($<0.5^\circ$ in $\lambda_\text{max}$ and $\phi_\text{max}$) and a small prediction horizon ($\lessapprox 14$hrs) causes a dramatic increase in the annual $\Delta v$ required. This is hypothesized to be due to the limit cycle due to radial perturbations at a stable longitude (see Figure~\ref{fig:radial_ignored}). With a short prediction horizon, the MPC optimization problem is not able to `see' far enough into the future to gather sufficient information on the expected motion. For smaller station-keeping windows, the satellite is hitting the boundary several times, often with great velocity, whereas larger window sizes allows the satellite to follow the drift limit cycle. Inclusions of velocity constraints may mitigate this anomaly, however this presents an opportunity for general caution. Prediction horizons less than half the length of an orbital period may be cause for an unexpected and significant increase in control input. Further, Figure~\ref{fig:Stable_sweep_partial} demonstrates that the station-keeping window size is the largest driver in decreasing $\Delta v$ for station keeping at stable longitudes.

\subsection{Epoch Sweep}
\begin{figure}[t!]
	\includegraphics[width=0.95\linewidth]{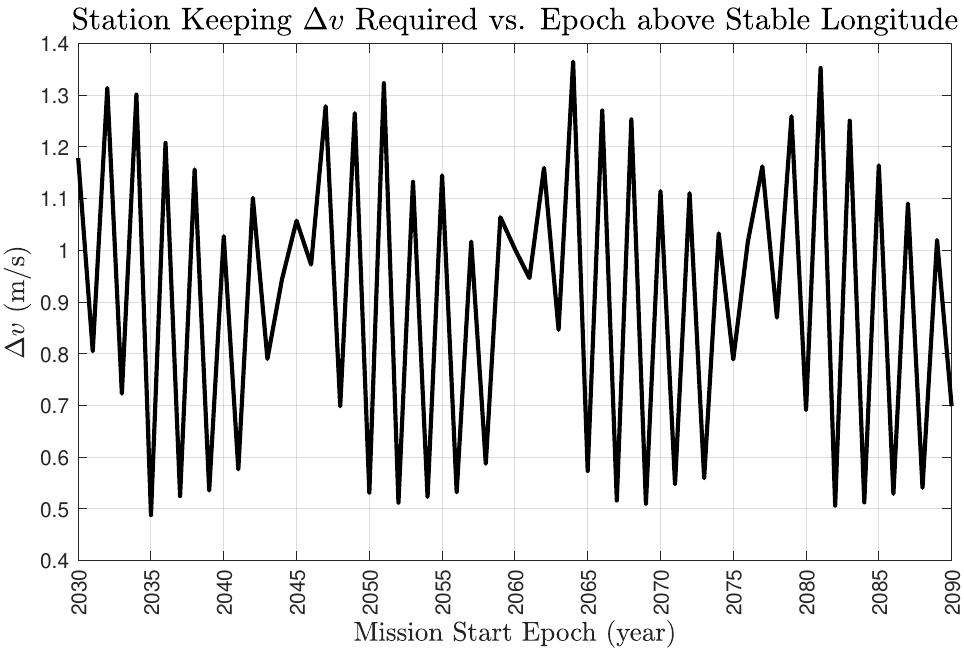} 
	\centering
	\caption{Station-keeping $\Delta v$ required vs. mission start date on January 1st at 12:00PM UTC at the year specified over the stable longitude of $17.92^\circ$~W. Simulations were performed for 668.6 orbits, or one Mars year.}\label{fig:stable_epoch}
\end{figure}

For a sweep of results across varying epoch, the MPC prediction horizon is chosen to be $N=16$~hours due to the diminishing returns of increased computation time and marginally lower $\Delta v$ with a longer $N$. The station-keeping window size is constrained to be $\lambda_\text{max} = \phi_\text{max} = 0.25^\circ$ to take advantage of significantly lower $\Delta v$ while constraining the maximum drift of the satellite to be reasonable. As was for the mission design in Section~\ref{sec:SMP}, larger station-keeping window size may cause complications in communications from the surface with minimal fuel performance gain. A sweep of simulations over a stable longitude with mission start dates varying one year between January 1st 2030 and 2090 using these parameters is shown in Figure~\ref{fig:stable_epoch}. 

In this epoch sweep, the largest annual $\Delta v$ required is at a mission start epoch in the year 2064 at $1.36495$~m/s whereas the smallest annual $\Delta v$ is in 2035 at $0.488062$~m/s. That is a factor of $\approx 2.8$ between the minimum and maximum. This large difference is due to the perturbations which are driving the drift of the satellite at a stable longitude---most of which are epoch dependent regarding the location of celestial bodies. There is again a cyclic nature in the $\Delta v$ required vs. epoch start date, where alternating years can vary drastically with a cycle of approximately 15 years where the annual difference is minimized. As mission length increases, this result is expected to smooth out through more time and fuel accumulation. However, for a shorter expected mission length, one might consider mission start dates with lower required $\Delta v$ for significant fuel savings. 

\begin{figure}[t!]
	\begin{subfigure}[]{1\textwidth}
        \centering
		\includegraphics[scale=0.65]{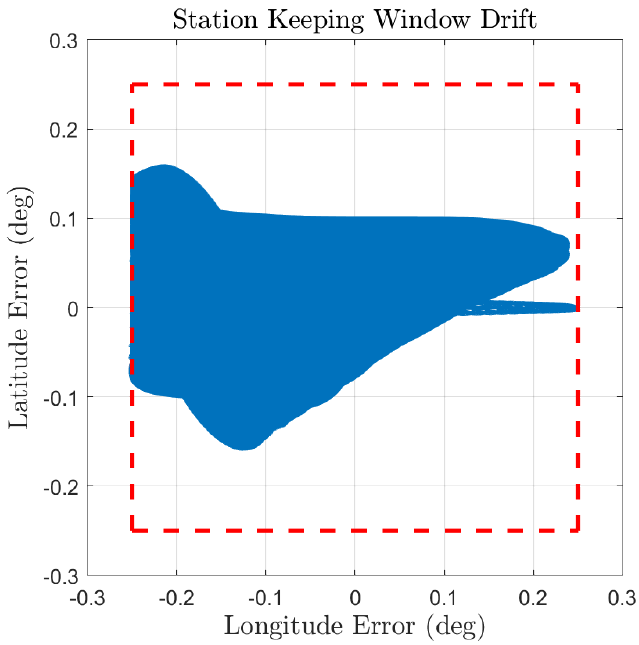} 
		\caption{}\label{fig:SMP_Window}
	\end{subfigure} \\
	\begin{subfigure}[]{1\textwidth}
        \centering
		\includegraphics[scale=0.65]{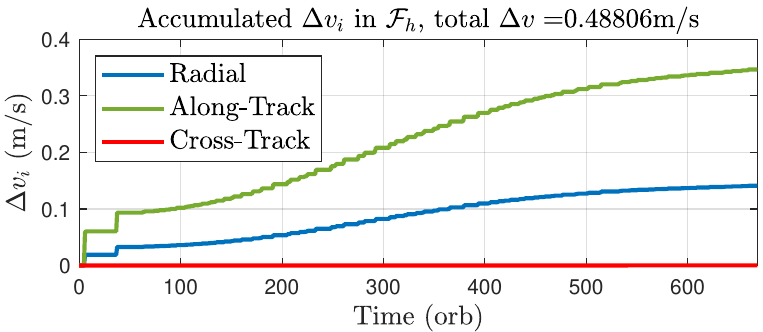} 
		\caption{}\label{fig:SMP_DV}
	\end{subfigure} \\
    \begin{subfigure}[]{1\textwidth}
        \centering
		\includegraphics[scale=0.65]{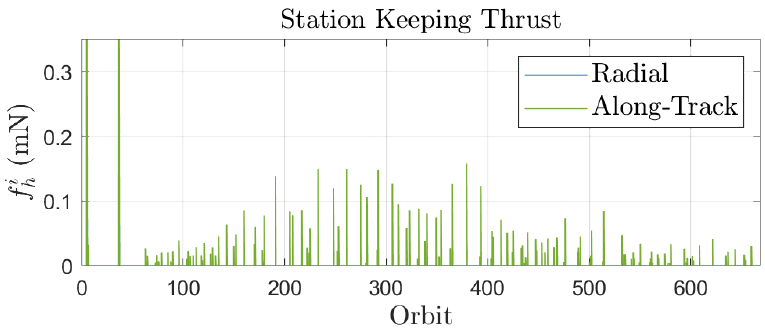} 
		\caption{}\label{fig:SMP_Thrust}
	\end{subfigure}%
	\caption{Station-keeping results over the stable longitude of $17.92^\circ$~W, with (a) the satellite trajectory (blue) within the station-keeping window (red), (b) accumulated $\Delta v$ over 668.6 orbits, and (c) a time history of radial and along-track thrust used. }\label{fig:stable_expanded}
\end{figure}

Results from the epoch start date with the smallest $\Delta v$ requirement at 2035 are shown in Figure~\ref{fig:stable_expanded}. The time-history of the satellite within the station-keeping window shows the satellite is capable of utilizing much more drift given that the disturbances at a stable longitude are much smaller in magnitude than the unstable longitude counterpart. It is expected that as the mission duration lengthens, the satellite will reach the North-South boundary of the station-keeping window, which in turn will increase the $\Delta v$ to counteract the inclination evolution of the satellite. The accumulation of $\Delta v$ shows primarily along-track thrust being utilized with an occasional radial thrust, which is much smaller in magnitude (also visible in the time history of thrust used in Figure~\ref{fig:SMP_Thrust}). The low magnitude of thrust and opportunity for on-off thruster pulses makes this implementation at a stable longitude a great candidate for electric propulsion.

The $\Delta v$ required for station keeping of a satellite placed directly over the Southern Meridiani Planum decreases significantly from approximately 28~m/s to less than 1~m/s when a satellite is placed at the nearest stable equilibrium longitude. Importantly, this satellite can still support communication with a surface station. It is clear that predictive control is a viable option for autonomous control of satellites in a stable areostationary orbit, especially with the various controller tuning capabilities and the careful placement of satellites over or near equilibrium longitudes. 

\section{Conclusions}
This paper explored the application of nonlinear model predictive control techniques to the autonomous station-keeping control of satellites in areostationary orbit around Mars. Through analysis via a high-fidelity satellite and environment model, the controller was verified and its performance across longitudes was compared to station-keeping $\Delta v$ estimates found via qualitative analysis existing in the literature. It was found that the predictive controller can nearly match the best-case estimate while taking into account \textit{all} important perturbations and satellite motion in simulation. 

A sample mission design for an areostationary satellite that could support ground operations at the Southern Meridiani Planum---a site for potential future crewed landings---was explored. The controller was appropriately tuned through several simulations by varying important prediction horizon and station-keeping window parameters in order to reduce the Mars-annual $\Delta v$ required for station keeping. A sweep of mission start epoch was completed using the appropriate controller parameters and constraints, where it was found that mission epoch has a negligible effect on $\Delta v$ requirements. These initial results were challenged via a modified areostationary satellite mission by exploring the placement of the satellite over the nearest stable longitude instead, where massive improvements in the reduction of $\Delta v$ required for station keeping were identified. 

Future work will include exploring more computationally-efficient station-keeping policies, as current satellite hardware may not be capable of solving a nonlinear optimization problem in an appropriate amount of time. This would require careful linearization about the open-loop satellite trajectory due to radial disturbances alone, such that the nonlinear coupling of these perturbations are inherently taken into account for a linear-quadratic MPC optimization problem.

\appendix 

\section{Communication availability between AMO and Earth}

The geometry of AMO and Mars' tilt relative to the ecliptic plane makes it expected for an areostationary satellite to be visible from Earth for a majority of the time. Mars' inclination off the ecliptic plane of Earth is $1.8^\circ$, thus it is assumed that Mars orbits in the same ecliptic to simplify calculations. The inclination of AMO off of the ecliptic is $\alpha = 25.19^\circ$ according to this assumption, with a semi-major axis of $A = 20427.7$~km. Via the simple equation 
\begin{equation*}
    h_{max}~=~A\cdot\sin\left(\alpha\right),
\end{equation*}
the maximum `height' of the satellite above the ecliptic plane is expected to be $8694.5$~km, much farther above Mars, which has an average volumetric radius of $3389.5$~km. This means that when the Earth is in the direction perpendicular to the AMO's right-ascension of the ascending node (RAAN, relative to the ecliptic), the satellite will be continually visible. 

However, the worst-case scenario occurs when the relative positions of Earth and Mars are aligned such that Earth's position aligns with the RAAN, and Mars occludes the view of the satellite from Earth. The angle of the orbit swept out by Mars' occultation is solved via the equation
\begin{equation*}
    \theta = \arccos\left(\frac{2A^2 - (2R_\text{m})^2}{2A^2}\right),
\end{equation*}
which provides the worst-case angle of $\theta = 19.1^\circ$. Thus, the length of time that the satellite-Earth communication would be occluded is 1 hour, 18 1/2 minutes, or only 5.3\% of one complete orbit.

The terms used in these calculations are provided the Mars Fact Sheet from NASA Goddard in Ref.~\cite{williams2004mars}.

\section*{Acknowledgments}
R. D. Halverson acknowledges partial support by the University of Minnesota - Twin Cities Undergraduate Research Opportunities Program (UROP). Halverson also acknowledges partial support by the Science, Mathematics, and Research for Transformation (SMART) Scholarship-for-Service Program within the Department of Defense (DoD). 

Simulation data was acquired via the Minnesota Supercomputing Institute (MSI). The authors would like to acknowledge Nathan Gall with the University of Minnesota - Twin Cities for collecting much of the data from the MSI. Data collection by Nathan Gall was partially supported by the Office of the Vice President for Research, University of Minnesota.

\bibliographystyle{elsarticle-num} 
\bibliography{Bibliography}

\end{document}